\documentclass[journal]{IEEEtran}
\usepackage{amsmath,amsfonts}
\usepackage{algorithmic}
\usepackage{algorithm}
\usepackage{array}
\usepackage{url}
\usepackage{verbatim}
\usepackage{graphicx}
\usepackage{cite}
\usepackage{subfigure,color}
\newtheorem{remark}{Remark} 

\begin{document}

\title{Sparsity-Driven EEG Channel Selection for Brain-Assisted Speech Enhancement}

\author{Jie Zhang, Qingtian Xu, Zhenhua Ling, Haizhou Li
\thanks{This work is supported by the National Natural Science Foundation of China (62101523) and USTC Research Funds of the Double First-Class Initiative (YD2100002008). The associate editor coordinating the review of this manuscript and approving it for publication was Prof. ********.}
\thanks{Jie Zhang, Qingtian Xu and Zhenhua Ling are with  the National Engineering Research Center for Speech and Language Information Processing (NERC-SLIP), University of Science and Technology of China (USTC), 230026, Hefei, China.  (e-mail: jzhang6@ustc.edu.cn; qingtianxu@mail.ustc.edu.cn; zhling@ustc.edu.cn)}
\thanks{Haizhou Li is with the Guangdong Provincial Key Laboratory of Big Data
Computing, The Chinese University of Hong Kong (CUHK), Shenzhen 518172, China. (e-mail: haizhouli@cuhk.edu.cn)}
\thanks{Digital Object Identifier:}
}
\markboth{Sparsity-Driven EEG Channel Selection for Brain-Assisted Speech Enhancement}%
{Shell \MakeLowercase{\textit{et al.}}: A Sample Article Using IEEEtran.cls for IEEE Journals}

\maketitle

\begin{abstract}
Speech enhancement is widely used as a front-end to improve the speech quality in many audio systems, while it is hard to extract the target speech in multi-talker conditions without prior information on the speaker identity. It was shown that the auditory attention on the target speaker can be decoded from the electroencephalogram (EEG) of the listener implicitly. In this work, we therefore propose a novel end-to-end brain-assisted speech enhancement network (BASEN), which incorporates the listeners' EEG signals and adopts a temporal convolutional network together with a convolutional multi-layer cross attention module to fuse  EEG-audio features. Considering that an EEG cap with sparse channels exhibits multiple benefits and in practice many electrodes might contribute marginally, we further propose two channel selection methods, called residual Gumbel selection and convolutional regularization selection. They are dedicated to tackling  training instability and duplicated channel selections, respectively. Experimental results on a public dataset show the superiority of the proposed BASEN  over existing approaches.  The proposed channel selection methods can significantly reduce the amount of informative EEG channels with a negligible impact on the performance. 
\end{abstract}


\section{Introduction}
\IEEEPARstart{S}{peech} enhancement (SE) aims to separate the target speech signal from complex background interferences. In multi-talker conditions, the interference might include unattended speaker(s)~\cite{wang_supervised_2018}, where extraction of the target speaker in this case is also called the ``cocktail party" problem~\cite{cherry1953some}. The SE module is usually adopted as a necessary front-end in many applications, e.g., automatic speech recognition~\cite{2015Robust}, hearing aids~\cite{Yan2018A}, telephony, human-machine interaction, to improve the speech quality and speech intelligibility.

With the rapid development in deep learning, many models have been proposed for speech separation/enhancement, e.g., Conv-TasNet~\cite{luo2019conv}, DPRNN~\cite{luo2020dual}, TF-GridNet~\cite{tfgnet}. These methods usually estimate speaker-specific masks, which are then applied to the noisy mixture to recover the separated speech. In order to help extract the target speech, the target speaker's clue can be leveraged as reference, e.g., pre-enrolled speaker embedding~\cite{xu2020spex}, visual features~\cite{pan2022usev,xu2023multi,ephrat2018looking,wu2019time} (lip movement, body motion~\cite{pan2022speaker}). Although these extra modal inputs contain the speaker identity to some extent, while in reality it is difficult to acquire the acoustic and visual cues or pre-enrolled  embedding of the target speaker simultaneously.

Human auditory system is extremely efficient in extracting the attended speaker from multiple competing speakers in real-world talking scenarios. We can automatically switch the talking objective even at a very low loudness, i.e., {\it cocktail-party effect}. Recently, it was shown that the target speaker can be decoded from brain waves e.g., electroencephalogram (EEG) of listeners~\cite{mesgarani2012selective,o2017neural,aroudi2020cognitive}. This facilitates using the cortical measurements to assist target speaker extraction. Compared to other modalities, 
utilizing the listener's EEG signals as the assistive clue might be more friendly convenient, as generally the EEG cap for brain signal acquisition is more naturally integrated with acoustic devices, e.g., binaural hearing aids~\cite{aroudi2020cognitive}. 

In order to identify the target speaker, one can  reconstruct the target speech envelope using the EEG signals~\cite{o2015attentional,biesmans2016auditory}, which is sometimes called stimulus reconstruction in literature. The reconstructed speech envelope can then be used to distinguish the attended speaker by comparing the similarity between candidate speech envelopes. One major problem therein is the latency due to the dependence on a long decision window~\cite{geirnaert2021electroencephalography}. Nevertheless, speech envelope recovery provides a projection from the brain to the acoustic domain, which can help analyze how the brain perceives multiple speakers. 
As the speech envelope somehow contains the attention information, it is then natural to design a multi-modal speech separation model using EEG and acoustic signals. For example, in~\cite{van2016eeg,o2017neural,pu2019joint}, the speech envelope is first estimated from the EEG signals, multiple candidate speech signals are then estimated using a separation network, and finally the estimated envelope is compared with the estimated sources to find out the most matched one. This clearly follows a cascade pipeline, leading to a high latency. Note that for speech enhancement we are only interested in the target speech and separating all existing sources is completely unnecessary.

In \cite{ceolini2020brain}, a learning-based  brain-inspired speech separation (BISS) model was proposed, which extracts the envelope of the attended speaker from the EEG signals of the listener and then uses the envelope to perform joint speech extraction and separation. As the speech envelope reconstruction and speaker extraction modules are optimized independently in a non-end-to-end manner,  the resulting solution might be  sub-optimal. End-to-end brain-enhanced speech denoiser (BESD)~\cite{hosseini2021speaker} and U-shaped BESD (UBESD)~\cite{hosseini_end--end_2022} were thus proposed, which utilize feature-wise linear modulation (FiLM)~\cite{perez2018film} to extract neural feature. However, these methods fuse the binary modalities via simple concatenation, and cross-modal complementary feature is thus insufficiently leveraged.  In~\cite{qiu2023tf}, a data augmentation based  TF-NSSE method was proposed, which outperforms UBESD and alleviates the issue of insufficient paired EEG-speech data for speaker extraction. More recently, NeuroHeed was proposed in~\cite{pan2023neuroheed} by incorporating the temporal association between the EEG signals and the attended speech, which can achieve a better performance at the cost of a higher model complexity (and thus unsuitable in the case of attention switch across two speakers in an online mode).

These brain-assisted methods reveal that using EEG signals is helpful for SE, while in practice the large amount of EEG electrodes for data collection would cause a high hardware cost and setup time. It is also inconvenient to design wearable devices with many electrodes distributed on the entire EEG cap. This  inspires a growing desire for sparse electrode placement via channel selection~\cite{mirkovic2015decoding}. It can also prevent overfitting, decrease the computational load and improve the interpretability of the model by removing irrelevant channels~\cite{strypsteen2021end, narayanan2019analysis,narayanan2020optimal},  where a topology-constraint on electrodes can then be taken into account. In principle, we can treat channel optimization as group feature selection, but most existing feature selection methods are wrapper-based using a certain heuristic search on the space of all  feature subsets~\cite{liu2005toward,alotaiby2015review}. Models  have to be trained on every candidate subset to choose the optimal one. However, in the era of deep neural networks (DNNs), the combination of  pre-wrapper based channel selection and back-end SE modules would be rather time-consuming. Similarly, the application of statistical sensor selection methods~\cite{zhang2022frequency} would suffer from the same problem.

To tackle this issue,  embedded approaches are more preferable, allowing to jointly train the SE model and selection layer in a complete end-to-end fashion. In \cite{strypsteen2021end}, an end-to-end learnable EEG channel selection method was proposed using the Gumbel-softmax function~\cite{jang2016categorical}. This Gumbel channel selection (GCS) uses the Gumbel-softmax function to construct continuous approximations for  discrete selection variables, forming a concrete selector layer~\cite{abid1901concrete,singh2023fsnet}. To avoid selecting duplicated channels, a regularization function was then proposed in \cite{strypsteen2021end}. The GCS can achieve competitive results compared to state-of-the-art (SOTA) methods for motor execution and auditory attention detection (AAD) tasks. However, the training in the context of EEG-based SE is more difficult, because in case the temperature parameter is large, the selection weights sampled from the Gumbel-softmax distribution are almost uniformly distributed~\cite{jang2016categorical}. As the training of GCS  starts with a large temperature (though gradually decreases)~\cite{strypsteen2021end}, the selection layer cannot produce valid EEG inputs for the relatively larger SE module at the beginning, and the entire model will thus suffer from training inefficiency and instability. Moreover, the duplicated channel issue still exists and is even more serious when the amount of total channels is large.

The focus of this work is thus on the sparsity-driven EEG channel selection based brain-assisted SE task.  The contribution of this paper is twofold as follows. First, we build an end-to-end time-domain baseline, called brain-assisted SE network (BASEN), which is based on the typical Conv-TasNet~\cite{luo2019conv}. We use two encoders to extract  speech and EEG embeddings, respectively from the speech mixture and EEG data. In order to explore the complementary information, we design a convolutional multi-layer cross attention (CMCA) module to deeply fuse the embeddings, followed by a decoder to recover the target speech waveform. It is shown that under the same condition the BASEN  outperforms UBESD~\cite{hosseini_end--end_2022}, indicating the superiority of the proposed feature fusion approach.

Second, based on BASEN, we then propose two novel  end-to-end channel selection methods. The first one aims to address the training instability of GCS~\cite{jang2016categorical} in the context of SE, called residual Gumbel selection (ResGS). The proposed ResGS exploits residual connections and designs a weighted residual training structure, which are shown to be useful to improve the training efficiency. The entire training procedure is split into two stages: 1) adding a weighted residual connection~\cite{he2016deep} to the original GCS layer and padding the GCS output to the same size as the residual output in order to provide valid EEG features at the beginning of the training; and 2) fine-tuning to better fit the post-SE module. 
The second approach, called convolutional regularization selection (ConvRS),  aims to resolve the duplicated selections without a preset number of selected channels, which is a channel-wise convolutional self attention selection layer consisting of 1D depth-wise separable convolutions and a linear output layer to produce the selection vector. We also design two special regularization functions to constrain the selection vector, where one for the discreteness and the other to control the sparsity.  Results show that both selection methods outperform the SOTA GCS method~\cite{strypsteen2021end}, ResGS successfully addresses the training difficulty and ConvRS can remove the duplicated solutions. 

This work is built on the success of our conference precursor~\cite{zhang2023basen}, which only presents the initial BASEN model. Compared to~\cite{zhang2023basen}, the novel contributions of this work are summarized as follows: 1) the BASEN is taken as the baseline and more results are presented; 2) we propose the ResGS to tackle the training instability in case of combining the front-end channel selection and back-end SE modules;  and 3) we propose ConvRS  to solve the duplicated decisions of EEG channels. As such, the proposed sparsity-driven EEG-based SE model can exploit a unique EEG channel subset to combine with the audio modality. It is shown that using the selected channel subsets for BASEN will not evidently decrease the SE performance  (sometimes can be even better), implying that many EEG channels are rather noisy or contribute marginally in practice. To our knowledge, this work is the first attempt of incorporating channel selection for EEG-based SE.

The remainder of this paper is structured as follows. Sec.~\ref{sec:method} presents a detailed description of  BASEN and ResGS. Sec.~\ref{sec:ConvRS} describes the proposed ConvRS. In Sec.~\ref{sec:exp_setup}, we introduce the experimental setup, followed by results  in Sec.~\ref{sec:result}. Finally, Sec.~\ref{sec:conclusion} concludes and discusses limitations of this work.

\begin{figure*}[!t]
  \centering
    \includegraphics[width=1\textwidth]{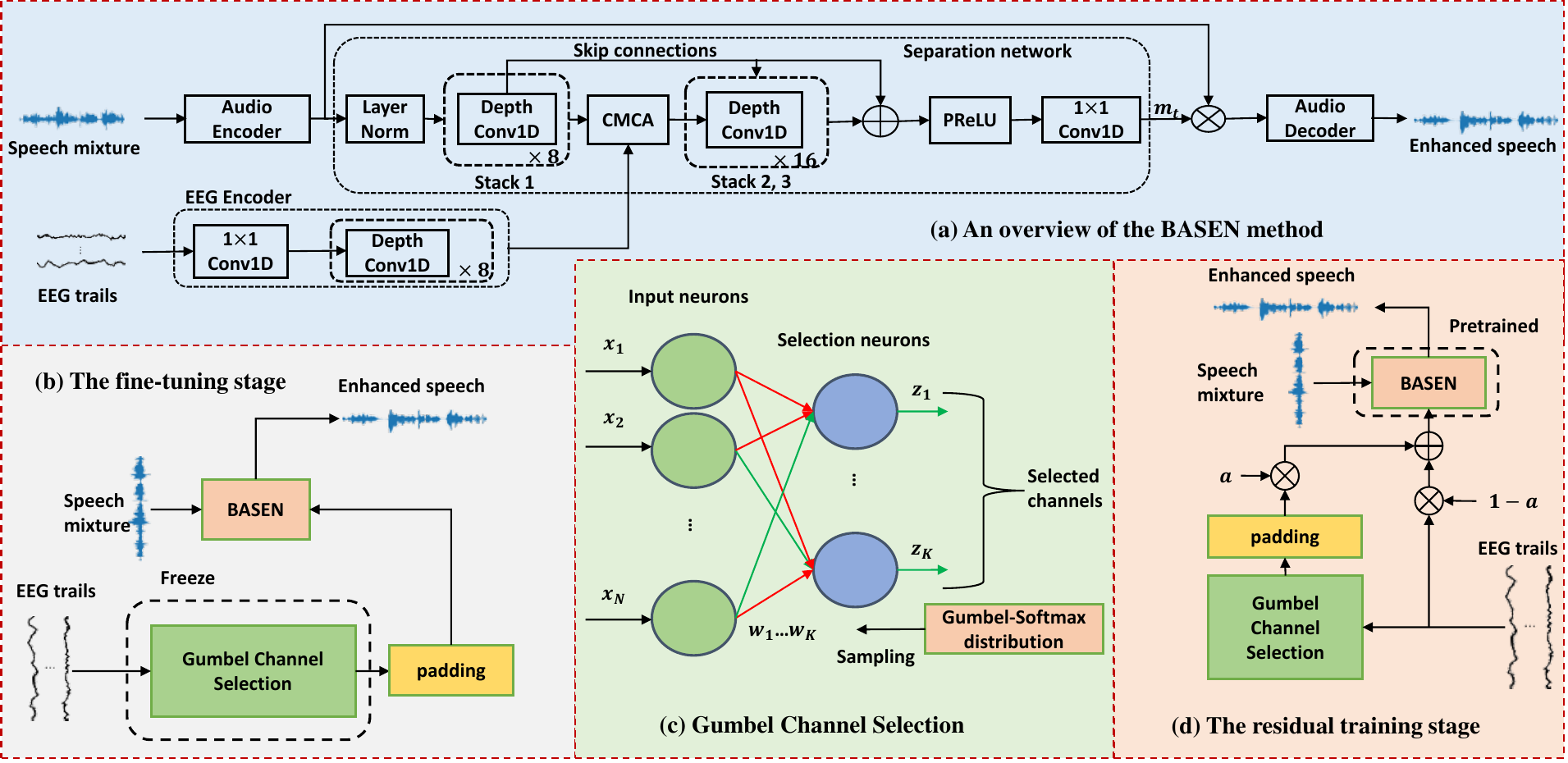}
  \caption{The proposed  ResGS-based sparsity-driven brain-assisted speech enhancement model: (a) the backbone BASEN model,  (b) ResGS fine-tuning stage, (c) Gumbel channel selection (GCS) and (d) ResGS residual training stage.}
  \label{fig:model}
\end{figure*}

\section{Sparsity-Driven BASEN}\label{sec:method}
In this section, we will introduce the proposed sparsity-driven BASEN model, including the baseline and ResGS-based EEG channel selection.
\subsection{Baseline model: BASEN}
As shown in Fig.~\ref{fig:model}(a), the proposed BASEN is a completely end-to-end time-domain speaker extraction model, which is built upon the typical Conv-TasNet~\cite{luo2019conv} backbone and adopts a temporal convolutional network (TCN) for both the EEG  and audio encoders. The CMCA module depending on the attention mechanism~\cite{vaswani2017attention} is designed  to deeply fuse the EEG and audio embeddings.

Given the noisy speech signal $x$ and the matched EEG signal $e_t$ recorded at the listener side, the audio encoder extracts the speech embedding sequence  $w_{x}$ from $x$, and the EEG encoder estimates the EEG embedding from the recorded EEG trials. That is,
\begin{equation}
    w_{x}={\rm AudioEncoder}(x), \quad
    e_{x}={\rm EEGencoder}(e_{t}).
\end{equation}
Both audio and EEG embeddings are input to the separation network to estimate the source-specific masks, given by
\begin{equation}
    [m_{0},\cdots,m_{T-1}]={\rm Separator}(w_{x},e_x),
\end{equation}
where $T$ denotes the number of existing sources. For target speaker extraction, $T$ = 2, i.e., the target speech and non-target components, while for the general speech separation task $T$ might be greater than 2. Note that the audio and EEG features have to be fused in the separator using the proposed CMCA module. Finally, the reconstructed  speech signal is obtained by mapping the audio embedding using the source-specific masks, given by
\begin{equation}
    \hat{s}_t={\rm Decoder}(w_{x}\odot m_t), \forall t\in\{1,\ldots,T\},
\end{equation}
where $\odot$ denotes the element-wise multiplication\footnote{The original Conv-TasNet was built for speech separation, which has to know the number of existing sources as {\it a priori}. For the considered target speaker extraction task, $T$ = 2 by default.}. 

\begin{figure}[!t]
  \centering  
    \includegraphics[width=0.3\textwidth]{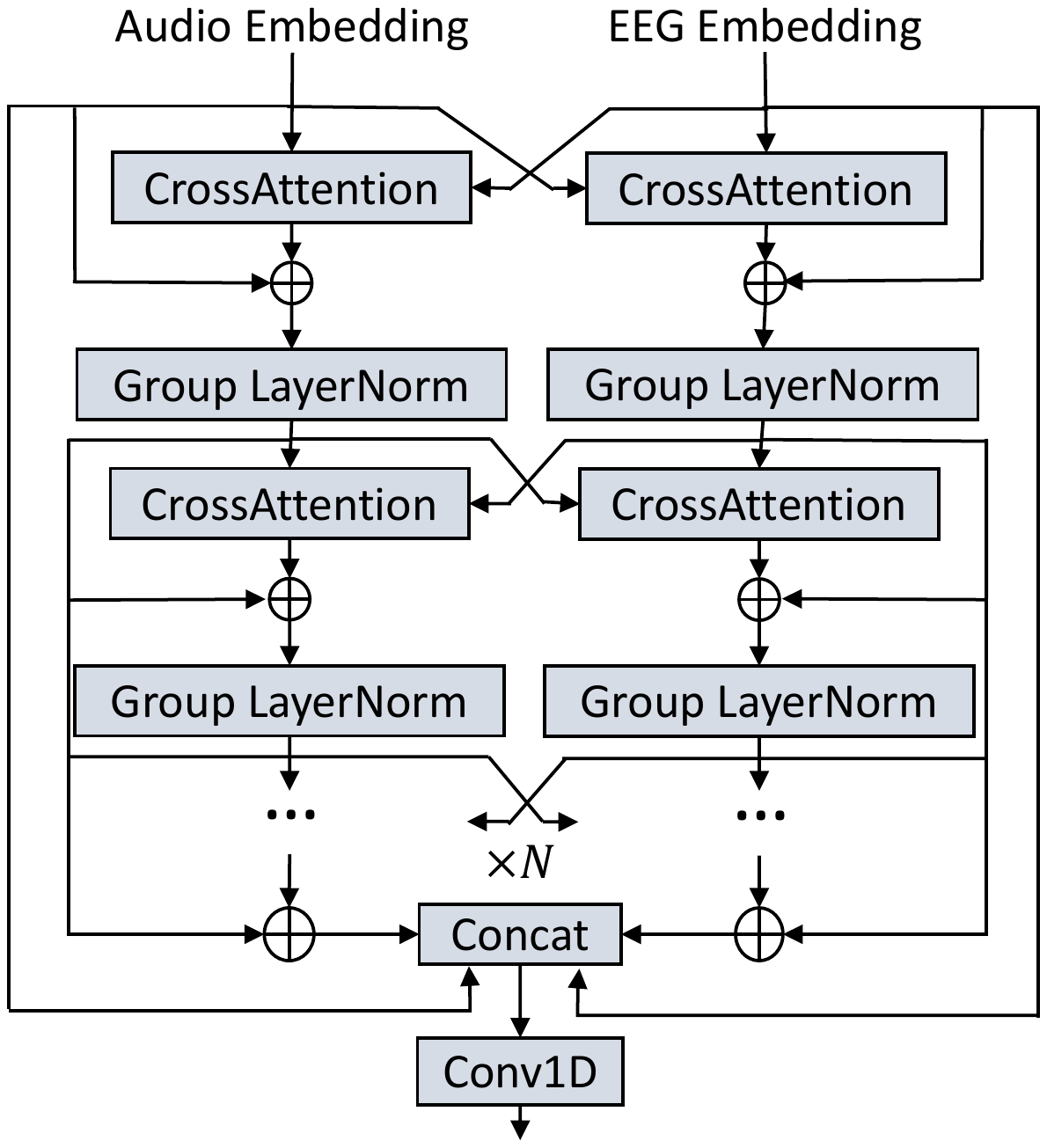}    
  \caption{The convolutional multi-layer cross attention (CMCA) module.}
  \label{fig:cmca}
\end{figure}

The audio encoder consists of a couple of 1-dimensional (1D) convolutions to extract speech embeddings and downsample the features. The corresponding audio decoder exhibits a mirror structure of the audio encoder, which has the same number of 1D transposed convolutions with a stride of 8 to perform upsampling. The EEG encoder is composed of one 1D convolution with a stride 8 to downsample the EEG trials and a stack of Depth-wise 1D convolutions proposed in~\cite{luo2019conv}, which has 8 layers and each  has residual connections to help form multi-level features. The Depth-wise convolution was shown to be effective in many tasks~\cite{kaiser2017depthwise,chollet2017xception}.

The CMCA module is designed for deep feature fusion, which is shown in Fig.~\ref{fig:cmca} consisting of $N$ layers of coupled cross attention blocks with skip connections and group normalization being placed between two adjacent layers. The left branch of CMCA deals with the audio stream and the right branch copes with the EEG modality. In layer $i$, the inputs of two branches are denoted by $e_{i-1}$ and $w_{i-1}$, respectively. Note that in case $i$ = 0, $e_0=e_x$ and $w_0=w_x$. Let the right cross attention block be denoted as ${\rm CrossAtt}_r$ and the left block as  ${\rm CrossAtt}_{l}$, respectively. Similarly, we define ${\rm GroupN}_l$ and ${\rm GroupN}_r$ as group normalizations. The outputs of layer $i,\forall i\ge 1$ can be represented as
\begin{align}
    w_{i}&={\rm GroupN}_{l}(w_{i-1}+{\rm CrossAtt}_{l}(e_{i-1},w_{i-1},w_{i-1})),\\
    e_{i}&={\rm GroupN}_{r}(e_{i-1}+{\rm CrossAtt}_{r}(w_{i-1},e_{i-1},e_{i-1})).
\end{align}
The audio-related and EEG-related layer-wise features of all layers are then added together, followed by concatenation  with the original audio-EEG embeddings $(w_x,e_x)$ over the channel dimension. The concatenated features will be sent into a 1D convolution to construct the fused dual-modal feature. 

\begin{figure*}[!t]
  \centering
    \includegraphics[width=0.98\textwidth]{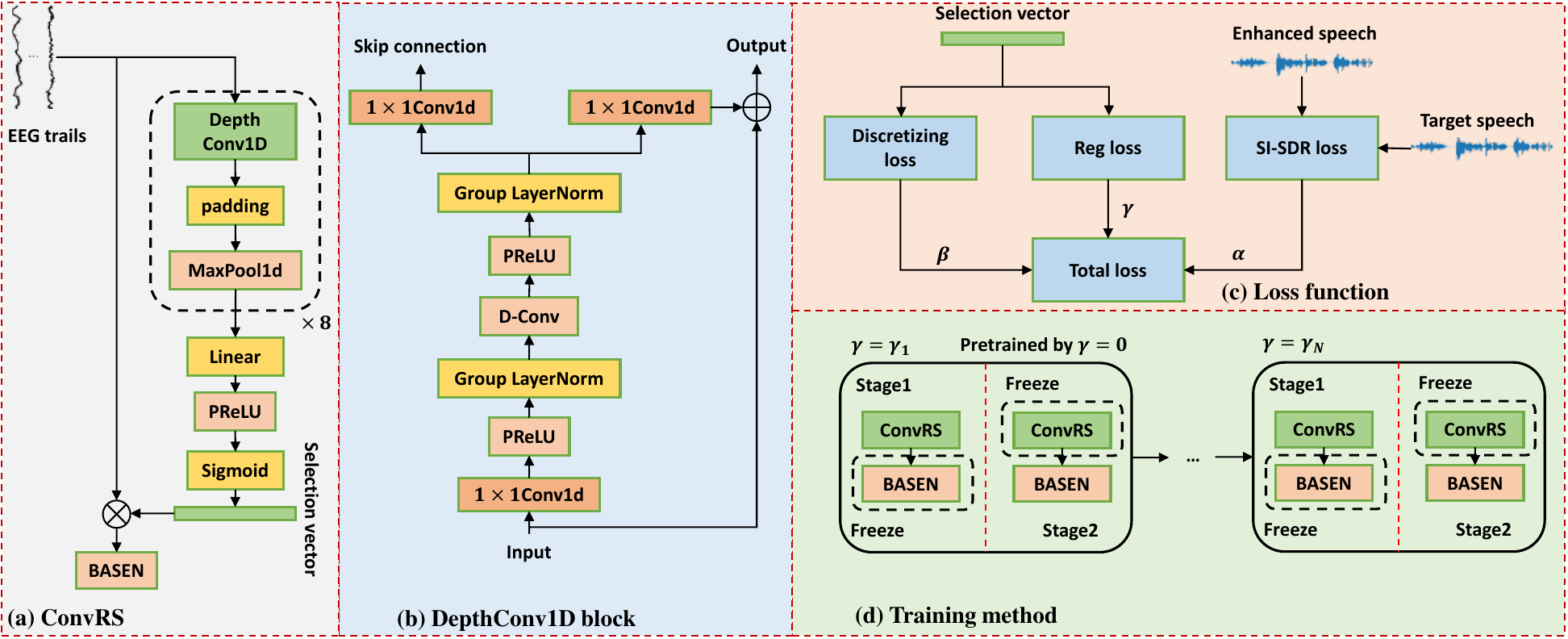}
  \caption{The proposed ConvRS-based sparsity-driven BASEN: (a) The ConvRS model,  (b) DepthConv1D, (c) Loss function and (d) The training process.}
  \label{fig:model2}
\end{figure*}

\subsection{ResGS}
The proposed ResGS method for sparse EEG channel selection can be regarded as an improved version of the original GCS~\cite{strypsteen2021end}, where the channel selection and SE modules are integrated in a fully end-to-end fashion in order to improve the training stability.

\textbf{Gumbel channel selection (GCS):} The GCS~\cite{strypsteen2021end} is the current mainstream end-to-end EEG channel selection method, e.g., see Fig.~\ref{fig:model}(c). Given $Q$ EEG channels, $K$ selection neurons are stacked on top of each other, where each indicates the selection or non-selection of the corresponding channel. Every selection neuron takes all EEG channels as input and produces a single output channel. Each neuron is parametrized by a learnable vector $\alpha_{k} \in {\rm I}\mathbb{R}_{>0}^{Q}$. In case the EEG input $e_{t}$ is fed into the selection layer, each neuron samples a weight vector $w_{k} \in {\rm I}\mathbb{R}^Q$ from the Gumbel softmax distribution (also called concrete distribution)~\cite{maddison2016concrete}, which is given by
\begin{equation}
    w_{nk}=\frac{{\rm exp}((\log\alpha_{nk}+G_{nk})/\tau)}{{\textstyle \sum_{j=1}^{Q}}{\rm exp}((\log\alpha_{jk}+G_{jk})/\tau) }, k \in \{1,...,K\},
\end{equation}
where $G_{nk}$ is sampled from the Gumbel distribution and $\tau \in (0, +\infty)$ is the temperature parameter. The output of the $k$-th neuron is then computed as $z_{k}=w_{k}^Te_{t}$.

In case $\tau$ approaches 0, the distribution turns to be more discrete and the sampled weights reduce to one-hot vectors approximately, so that each neuron can select a certain input channel~\cite{maddison2016concrete}. The probability $p_{nk}$ of neuron $k$ selecting  channel $n$ is given by
\begin{equation}
    p_{nk}=\frac{\alpha_{nk}}{{\textstyle \sum_{j=1}^{Q}}\alpha_{jk}}.
\end{equation}
The probability vectors $p_{k}=[p_{1k},\ldots,p_{Qk}]^T$ will converge to be one-hot as well. For testing, we can directly select the channel with the maximum probability at each neuron.

As the training of GCS usually starts from a large temperature $\tau$, which makes $w_{nk}$ close to $1/Q$~\cite{jang2016categorical}, the output of every selection neuron is almost the average of all channels. {The use of annealing strategy for training can lower the temperature and the resulting weights at different selection neurons might differ over time, but the outcomes heavily depend on downstream tasks. For small-size tasks, e.g., AAD, speech {envelope} construction~\cite{strypsteen2021end, narayanan2019analysis,narayanan2020optimal}, the front-end GCS can be easily trained. As the considered BASEN ($\approx$0.6M) is much larger than the selection layer, a direct joint training of front-end GCS and back-end BASEN is difficult, where many initial iterations seem irrelevant as the relatively larger SE module cannot obtain valid inputs from the selector layer.  The obtained selection weights cannot reflect informative task-guided clues, i.e., non-optimal in target speaker perception. Nevertheless, this classic GCS will still be compared in Sec.~\ref{sec:result}.}

\textbf{Residual training stage:} In order to alleviate the training difficulty in GCS, we design a special residual connection, see Fig.~\ref{fig:model}(d). Instead of directly using the output of the GCS $z=[z_{1},\ldots,z_{k}]^T$, we pad it along the channel dimension to be the same size of original EEG input, which are then added with the original EEG signal as
\begin{equation}\label{eq:padding}
    \bar{e}=(1-a)e_{t}+a{\rm Padding}(z),
\end{equation}
where $a$ is the weighting parameter (empirically set to be 0.1 in this work). This operation can guarantee a valid input for the BASEN, particularly when the temperature parameter is large at the beginning of training. 
The output $\bar{e}$ is then sent to the BASEN together with audio mixtures to perform target speaker extraction. Note that the backbone BASEN model\footnote{https://github.com/jzhangU/Basen.git} has to be pre-trained on the same dataset.

\textbf{Fine-tuning stage:} This stage, which is shown in Fig.~\ref{fig:model}(b), is designed to fine-tune the SE module. After the residual training, we remove the  residual connection and freeze the parameters in GCS. Given  the selection results obtained by GCS,  we set the selection layer to the test mode and fine-tune the BASEN using the selected input channels. As such, the training instability of GCS can be resolved in the case of choosing a channel subset for BASEN. Note that the proposed ResGS does not take the duplicated selections into account.
{
\begin{remark}
The proposed ResGS method involves a two-stage training procedure: 1) residual training to calculate channel-wise selection weights and 2) fine-tuning the pre-trained BASEN with the frozen GCS module to fit the SE task. The padding operation in (\ref{eq:padding}) is only considered in the first stage, which can help gradient back-propagation and stabilize the training. In the second stage only the  selected EEG channels are used to fine-tune the back-end BASEN. This two-stage design is more efficient than a straightforward combination of the classic GSC and back-end BASEN.
\end{remark}}

\section{Convolutional Regularization Selection}\label{sec:ConvRS}
In order to resolve the issue of duplicated selected channels in existing approaches and the proposed ResGS method and facilitate the EEG channel selection solution to be independent on the prescribed number of the selected channels, we further propose a ConvRS method by incorporating two regularization functions to constrain the selection vector in this section.

\subsection{Model configuration} The structure of the proposed ConvRS is shown in Fig.~\ref{fig:model2}(a), which consists of several special Conv1D blocks and a linear layer with a parametric ReLU function and a sigmoid function. Each block contains a DepthConv1D block, a padding layer and a 1D MaxPool layer. The DepthConv1D blocks are depth-wise separable Conv1D~\cite{luo2019conv}, which contains less parameters than normal Conv1D but obtains a comparable performance. As shown in Fig.~\ref{fig:model2}(b), the considered DepthConv1D can decouple the standard convolution operation into two consecutive operations, i.e., a Depth-wise convolution followed by a point-wise convolution. As such, the model size can be largely reduced. The MaxPooling layer is used to downsample the feature with  a stride of 2. The padding layer is to keep the feature length being the half of last feature length after each block. Therefore, the output of  the designed Conv1D block of the $i$-th layer can be written as
\begin{equation}
    e_{i}={\rm MaxPool1d}({\rm padding}({\rm DepthConv1D}(e_{i-1}))),
\end{equation}
where  $e_{0}=e_t$.  The selection vector is then given by the linear layer as
\begin{equation}
    \mathbf{s}={\rm Sigmoid}({\rm PReLU}({\rm Linear}(e_{N_{1}}))) \in [0,1]^Q,
\end{equation}
where  $N_{1}$ denotes the number of Conv1D blocks. Applying the selection vector, the sparse EEG signals can be given by
\begin{equation}
    e_{\rm ConvRS} = e_{t} \odot {\rm expand}(\mathbf{s}),
\end{equation}
where the $\rm expand$ function is to make the shapes of $\mathbf{s}$ and original EEG signals consistent.

\subsection{Regularization-based loss function} 
\begin{figure}[!t]
\centering
    \includegraphics[width=0.45\textwidth]{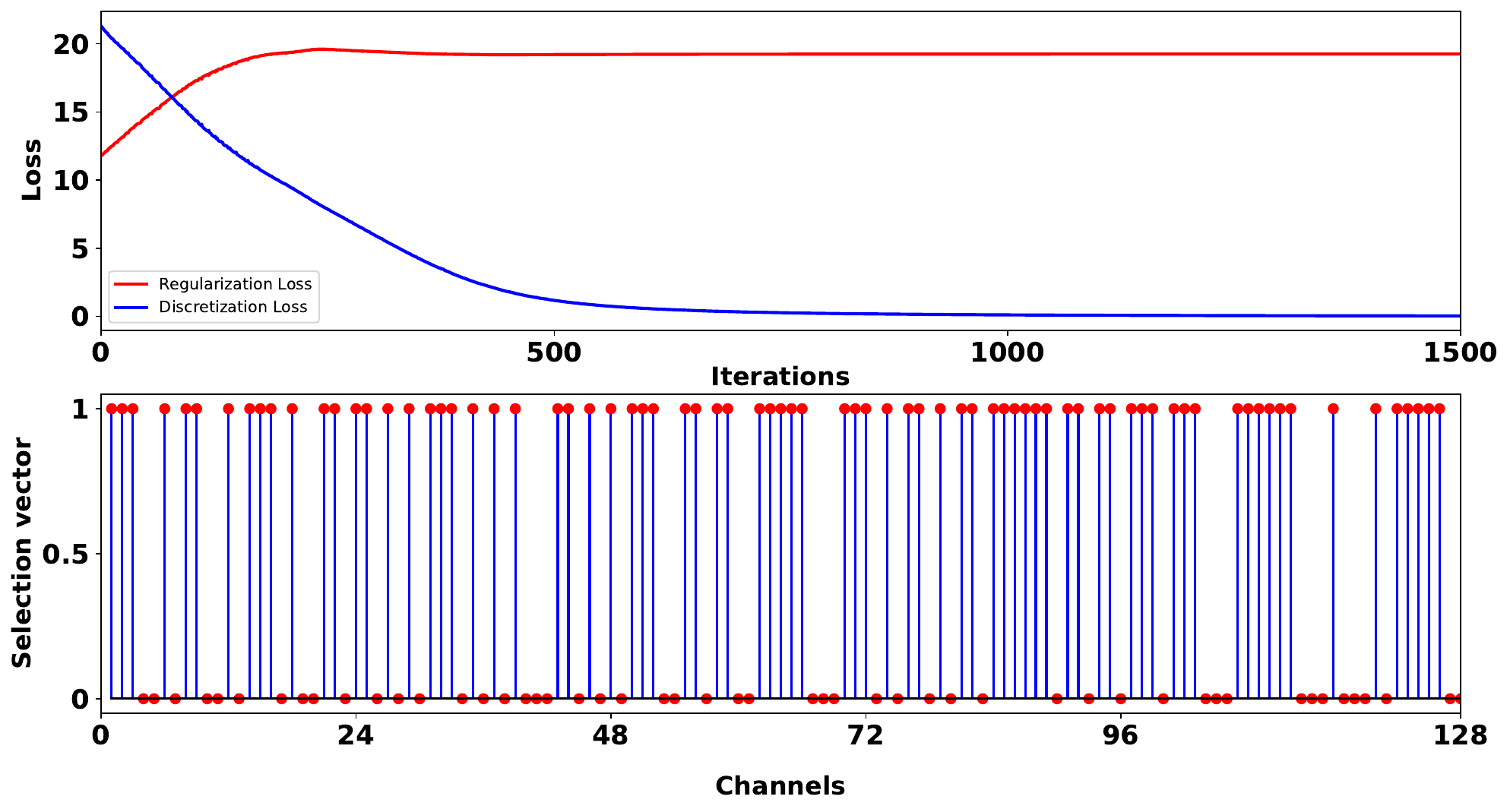}
  \caption{The loss function and final selection vector of ConvRS with $\gamma$ = 0.01.}
  \label{fig:loss}
\end{figure}

We consider two regularization strategies to accomplish the selection process and constrain the cardinality of the selected channel subset, which are depicted in Fig.~\ref{fig:model2}(c). The first is discretization loss, which constrains the discreteness of the selection vector. For each element in the selection vector, we treat it as the selection probability of the corresponding EEG channel. The discreteness loss function can be written as
\begin{equation}\label{eq:loss_Ld}
    \mathcal{L}_{d}=k_{1}\left(-\frac{\mathbf{d}^T \mathbf{d}}{QB} + b\right),
\end{equation}
where $B$ denotes the batch size, $k_{1}$ is the scaling coefficient to balance different loss components (which is empirically set to 100  in this study), and the distance vector is defined as
\begin{equation}
    \mathbf{d}=\mathbf{s}-q,
\end{equation}
where $q$ is set to 0.5 to diversify the selection probabilities, since the probability of 0.5 corresponds to the invalid decision of ``toss a coin". {Note that as $\mathbf{d}^T\mathbf{d}$ should be computed over all batches, leading to 0 $\leq \frac{\mathbf{d}^T \mathbf{d}}{QB}\leq$ 0.25 with $Q$ = 128 in experiments and the maxima given by any Boolean vector $\mathbf{s}$, we add a constant bias $b$ = 0.25 in (\ref{eq:loss_Ld}) to set the minima of $\mathcal{L}_d$ to be 0. As such, we can see that 25 $\ge \mathcal{L}_d\ge$ 0 with the minima reached by any Boolean vector $\mathbf{s}$, e.g., see Fig.~\ref{fig:loss}. This loss enforces the selection solution $\mathbf{s}$ to be discrete as a Boolean selection vector.}

In addition, to decrease the cardinality of the selected EEG channel subset that equals the $\ell_0$ norm of $\mathbf{s}$, we consider its convex relaxation version and incorporate the $\ell_2$ norm as the sparsity regularization term, i.e.,
\begin{equation}
    \mathcal{L}_{\rm reg}=k_{2}||\mathbf{s}||_2^2,
\end{equation}
where $k_{2}$ is a similar scaling balance parameter, which is set to 0.25 in this work. {Note that as $\ell_2$ norm is not sparsity-inducing in general, the sparsity regularization loss usually has to be combined with a posterior thresholding or rounding operation. However, the simultaneous incorporation of the discreteness loss $\mathcal{L}_d$ and the sparsity loss $\mathcal{L}_{\rm reg}$ can lead the raw selection vector $\mathbf{s}$ to be closely Boolean.  This is shown in Fig.~\ref{fig:loss}, where the selection vector is obtained without any thresholding using the experimental setup in Sec.~\ref{sec:exp_setup}. The final selection status can thus be simply resolved by rounding. The proposed loss functions are therefore valid to constrain channel selection.

}

The third component of the loss function is designed for target speaker extraction, which is chosen as the often-used scale-invariant signal-to-distortion ratio (SI-SDR)~\cite{le2019sdr} to measure the speech distortion, given by
\begin{equation}
    \mathcal{L}_{\rm SI\text{-}SDR}=10\log_{10}\frac{\left \| x_\text{target} \right \|^2  } {\left \| x_\text{res} \right \|^2  },
\end{equation}
where $x_\text{target}=\frac{\hat{s}_t^Ts}{\Vert s\Vert ^2}s$ and $x_\text{res}=x_\text{target}-\hat{s}_t$ with $s$ and $\hat{s}_t$ denoting the target speech and the reconstructed speech signals, respectively. In general, a higher SI-SDR means a better speech quality. The negative SI-SDR is thus taken as the loss function for training.  It was shown that SI-SDR is a well-performed general purpose for the time-domain SE~\cite{kolbaek2020loss}.

Altogether, the total loss function can be formulated as
\begin{equation}\label{eq:total-loss}
    \mathcal{L}=-\alpha \mathcal{L}_{\rm SI\text{-}SDR}+\beta \mathcal{L}_{d}+\gamma \mathcal{L}_{\rm reg},
\end{equation}
where $\alpha$, $\beta$ and $\gamma$ are weights of the three components. In this work, both $\alpha$ and $\beta$ are set to 0.5, but $\gamma$ is set depending on the expected amount of EEG channels, i.e., $\gamma$ controls the sparsity of the selected channel subset.

\subsection{Progressive training}\label{sec:progress_training}
In order to stabilize the joint training of the front-end ConvRS and back-end BASEN, particularly in case of a large $\gamma$-value, we propose a progressive training algorithm in this context, which is shown in Fig.~\ref{fig:model2}(d). In detail, we initially set $\gamma$ to 0 and train the whole model without any pre-training, resulting in the original selection output, which can also be viewed as a raw channel selection solution. For the training of other $\gamma$-values, we then use the trained model with $\gamma$ = 0 as the pre-trained model. Given this initialization, next we perform a two-stage training: 1) in the first stage we freeze the BASEN module and only train the ConvRS for a few epochs; and 2) in the second stage we  freeze the ConvRS module and fine-tune the BASEN module to adapt to the selected channels. Then for subsequent $\gamma$-values, the pre-training model is replaced with the last model with the latest $\gamma$-value. This progressive training process will be terminated until all $\gamma$-values are considered. Note that given a limited number of EEG channels $Q$, the candidate $\gamma$-set is also constrained.
{
\begin{remark}
Notice that the removal of the discreteness loss  $ \mathcal{L}_{d}$  from (\ref{eq:total-loss}) would make the proposed ConvRS method act as a plain attention mechanism, where the obtained selection vector $\mathbf{s}$ can be interpreted as the attention weights of $Q$ EEG channels. The considered loss function enables the selection to be both sparse and discrete, which is input-independent after training over all subjects at least on the dataset in Sec. \ref{sec:exp_setup}. It experimentally turns out that the sparsity only depends on the user-defined parameter $\gamma$.  We admit that this has to be validated in the case of a larger data diversity, and we will release a new dataset in this field soon. During the testing phase, we only include the channels indicated by $s_j$ = 1, $\forall j\in \{1,\ldots, Q\}$. Hence, the channel selection of ConvRS is also static, similarly to ResGS. The superiority of ConvRS over ResGS lies in the removal of duplicated channel selections owing to the single selector $\mathbf{s}$.
\end{remark}
}

\section{Experimental Setup}\label{sec:exp_setup}
\subsection{Dataset}
We evaluate the proposed methods using the dataset originated from \cite{broderick2018electrophysiological}. {All procedures of using this dataset for performance evaluation follow a strict adherence to the Declaration of Helsinki and the consent from the Ethics Committees of the School of Psychology at Trinity College Dublin and the Health Sciences Faculty at Trinity College Dublin. This dataset includes 33 subjects in total (28 males and 5 females) with a mean age of 27.3$\pm$3.2 years, and all subjects participated in experiments were native English speakers with a normal hearing level and no history of neurological diseases. Similarly to \cite{hosseini_end--end_2022}, we exclude the data from the sixth subject due to the poor recording quality.

All subjects conducted 30 trials and each trial lasts 60 seconds. The audio stimuli are two stories read by different male speakers. During each trial, one story was presented to the left ear and the other played on the right  side. The subjects were asked to pay attention to the story from either the left (17) or the right ear (16 + 1 excluded subject) side, resulting in two groups. After each trial, they were required to answer multiple-choice questions on each story to validate if their attention is really on the instructed listening. In order to preserve a complete storyline, for each trial the story began where the last trial ended. Subjects were asked to visually focus on a cross hair on the screen center in order to minimize other interfering electroneurographic signals caused by e.g., eye blinking, physical motions.

During experiments, the subjects wore an EEG cap with 128 channels (plus two mastoids),  i.e., providing $Q$ = 128 EEG channel measurements at a rate of 512 Hz using a BioSemi ActiveTwo system. The EEG data were further downsampled to 128 Hz for consistency with \cite{hosseini_end--end_2022}. The audio stimuli were played using Sennheiser HD650 headphones at a sampling frequency of 44.1 kHz. In order to avoid the impact of the audio intensity on the attention, the stimuli were normalized to have the same RMS level in terms of amplitude.}
More details on the experimental setting can be found in \cite{broderick2018electrophysiological}.

\subsection{Pre-processing}\label{sec:preprocessing}
The considered experimental setup keeps the same as that in \cite{hosseini_end--end_2022,zhang2023basen}. In order to reduce the computational complexity, we downsampled the sound data to a sampling rate of 14.7 kHz. The two stimuli were then equally added to synthesize the noisy mixture at a fixed signal-to-noise ratio (SNR) of 0 dB. The presence of extra background noise or reverberation was not taken into account. We divided the data into the following three groups: randomly choosing 5 trials from all subjects for testing, 2 trials for validation and all the remaining trials for training. For the training and validation sets, each trial was cut into 2-second segments. For the testing set, each 60-second trial was cut into 20-second segments. {This dataset split keeps strictly the same as UBESD in~\cite{hosseini_end--end_2022}.}

The EEG data were first pre-processed using a band-pass filter with the pass frequency band ranging from 0.1 Hz to 45 Hz. To identify channels with excessive noises, the standard deviation (SD) of each channel was compared to the SD of the surrounding channels and each channel was visually inspected. Channels with excessive noises were re-calculated by spline interpolation of the surrounding channels. The EEGs were re-referenced to the average of the mastoid channels to avoid introducing noise from the reference site. To remove artifacts caused by eye blinking and other muscle movements, we performed independent component analysis (ICA) using the EEGLAB toolkit~\cite{delorme2004eeglab}. For each subject, the trial that contains too much noise was excluded from experiments.

In general, EEG signals are noisy mixtures of underlying existing sources that are not necessarily related to the target speech stimuli. It was shown in \cite{hosseini_end--end_2022} that using the underlying neural activity that is more related to the speech stimuli for the separation network is more beneficial for the performance than directly using the EEG signals. We thus further processed the EEG data using a frequency-band coupling model~\cite{whittingstall2009frequency} to estimate the audio related neural activity from EEG signals, which can be represented by the cortical multi-unit neural activity (MUA) from EEG signals. In \cite{whittingstall2009frequency,moinnereau2020frequency}, the MUA was shown to be effective for estimating the neural activity in the visual and auditory systems, which is given by
\begin{equation}
    U(t)=a_{\gamma}\times P_{\gamma}(t) + a_{\delta}\times \angle\delta(t),
\end{equation}
where $P_{\gamma}(t)$ and $\angle\delta(t)$ are the amplitude of the gamma band and the phase of the delta band, respectively, and $a_{\gamma}$ and $a_{\delta}$ are the corresponding coefficients, both of which are set to be 0.5 in experiments.

\subsection{Evaluation details}
For performance evaluation, we use three objective metrics to quantify the overall quality of the enhanced speech signal, including SI-SDR (in dB)~\cite{le2019sdr}, perceptual evaluation of speech quality (PESQ)~\cite{rix2001perceptual} and short-time objective intelligibility (STOI)~\cite{taal2010short}. For the training of ConvRS, the Adam optimizer~\cite{kingma2014adam} is exploited with a momentum $\beta_{1}$ of 0.9 and a denominator momentum $\beta_{2}$ of 0.999. The regularization weight $\gamma$ originates from a pre-defined set of \{0, 0.05, 0.1, 0.15, 0.2, 0.25, 0.3, 0.35\} to obtain the corresponding selected EEG channel subsets. We train 5 epochs in the first stage and 60 epochs in the progressive training. We use the linear warmup following the cosine annealing learning rate schedule with a maximum learning rate of 0.005 in the first stage and 0.0002 in the second stage with a warmup ratio of 5$\%$. For the BASEN and ResGS, we also use the linear warmup following the cosine annealing learning rate schedule with a maximum learning rate of 0.0002 and a warmup ratio of 5$\%$. The model is trained for around 60 epochs with a batch size of $B$ = 8. In the ResGS, The temperature parameter $\tau$ is decayed from 10 to 0.1.

\section{Experimental Results}\label{sec:result}
In this section, we conduct several comparisons to validate the proposed sparsity-driven brain-assisted SE methods. 

\begin{figure}[!t]
\centering
  \subfigure[Ablation study]{ 
   \centering
    \includegraphics[width=0.223\textwidth]{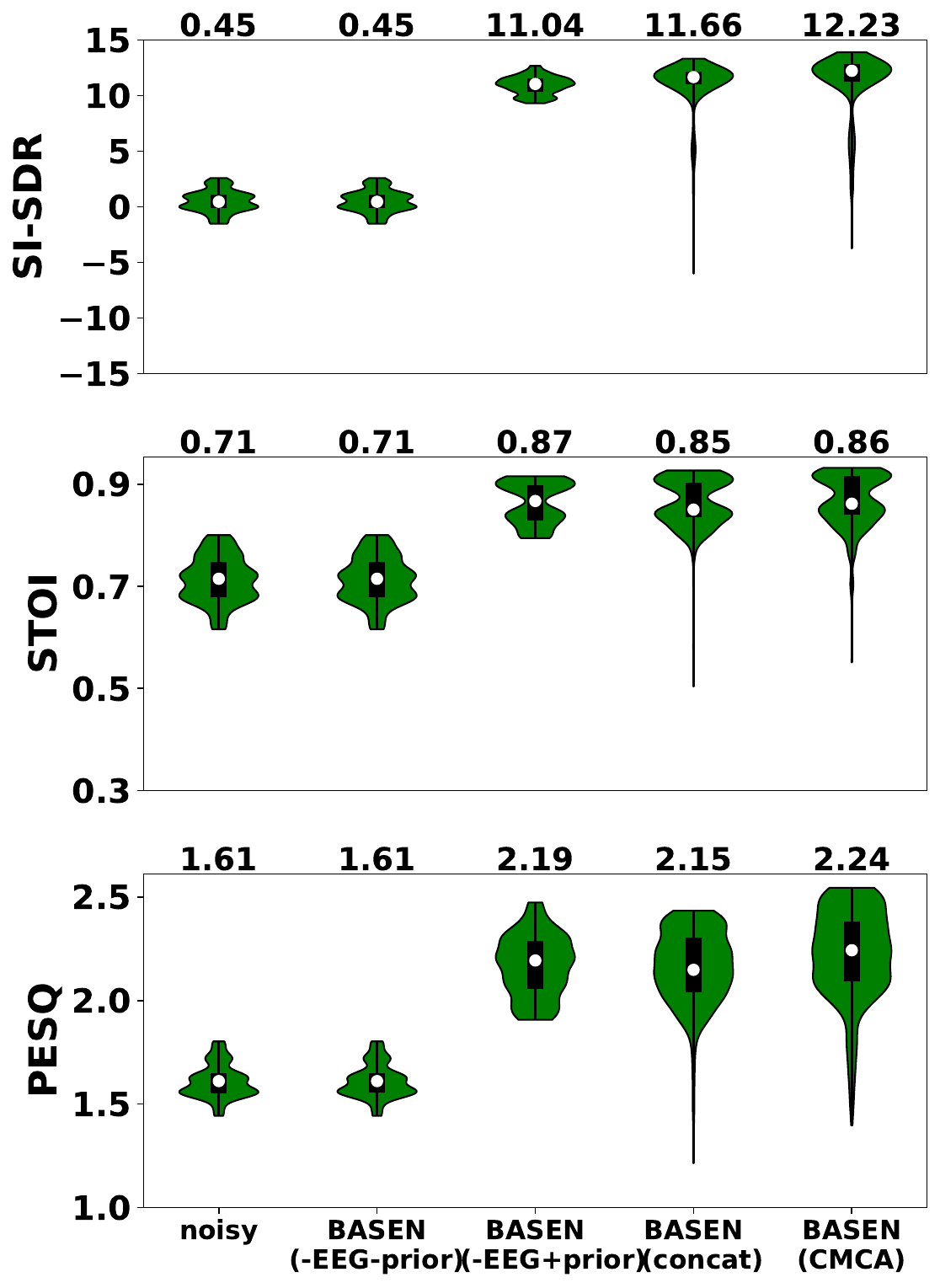} }
  \subfigure[Impact of CMCA layers]{
     \centering
    \includegraphics[width=0.223\textwidth]{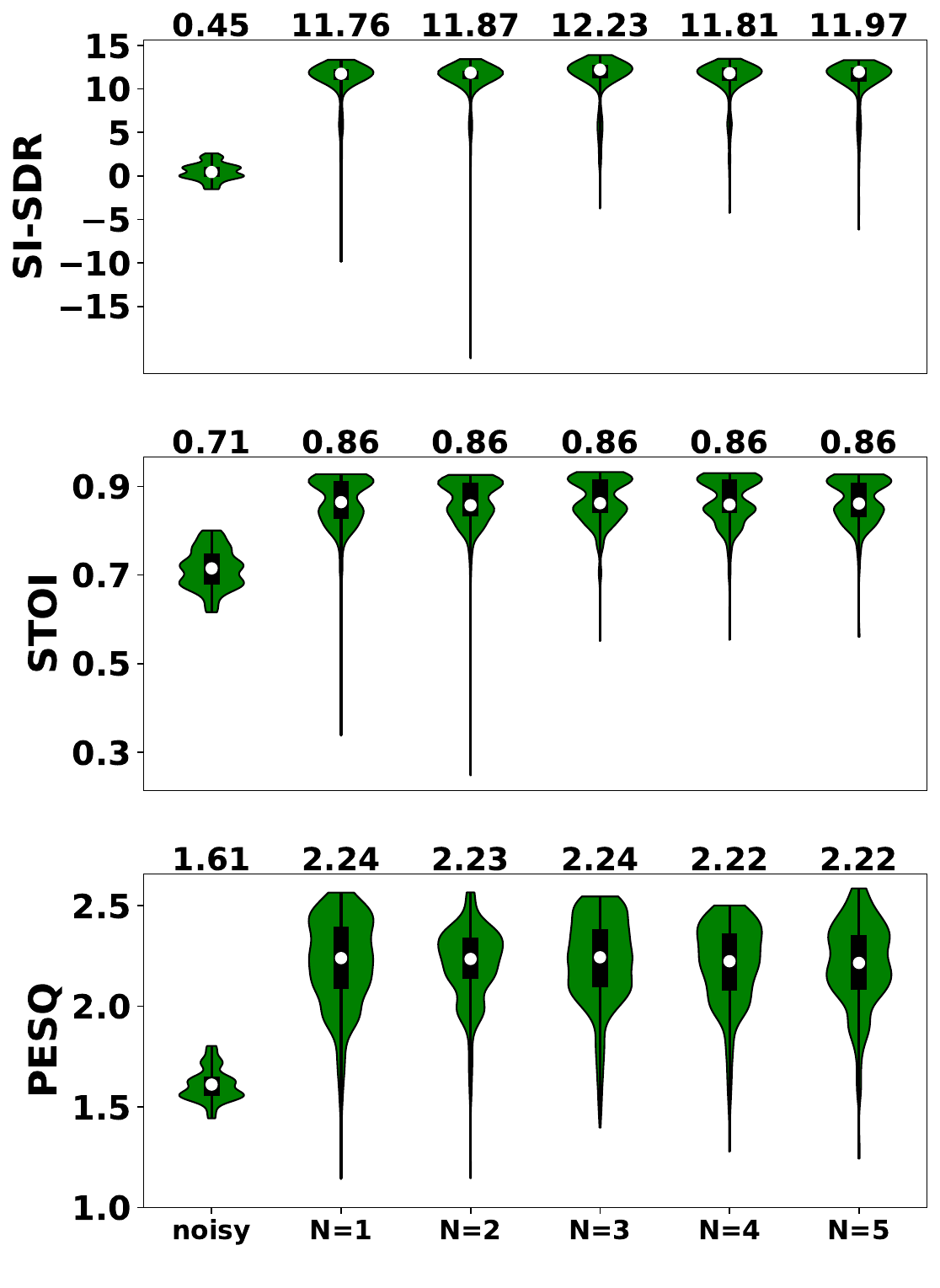}}
  \caption{The SE performance  for unknown attended speaker denoising in terms of SI-SDR , PESQ and STOI.}
  \label{fig:self-comparison}
\end{figure}

\begin{figure*}[!t]
  \centering
  \subfigure[UBESD~\cite{hosseini_end--end_2022}]{
  \centering
    \includegraphics[width=0.48\textwidth]{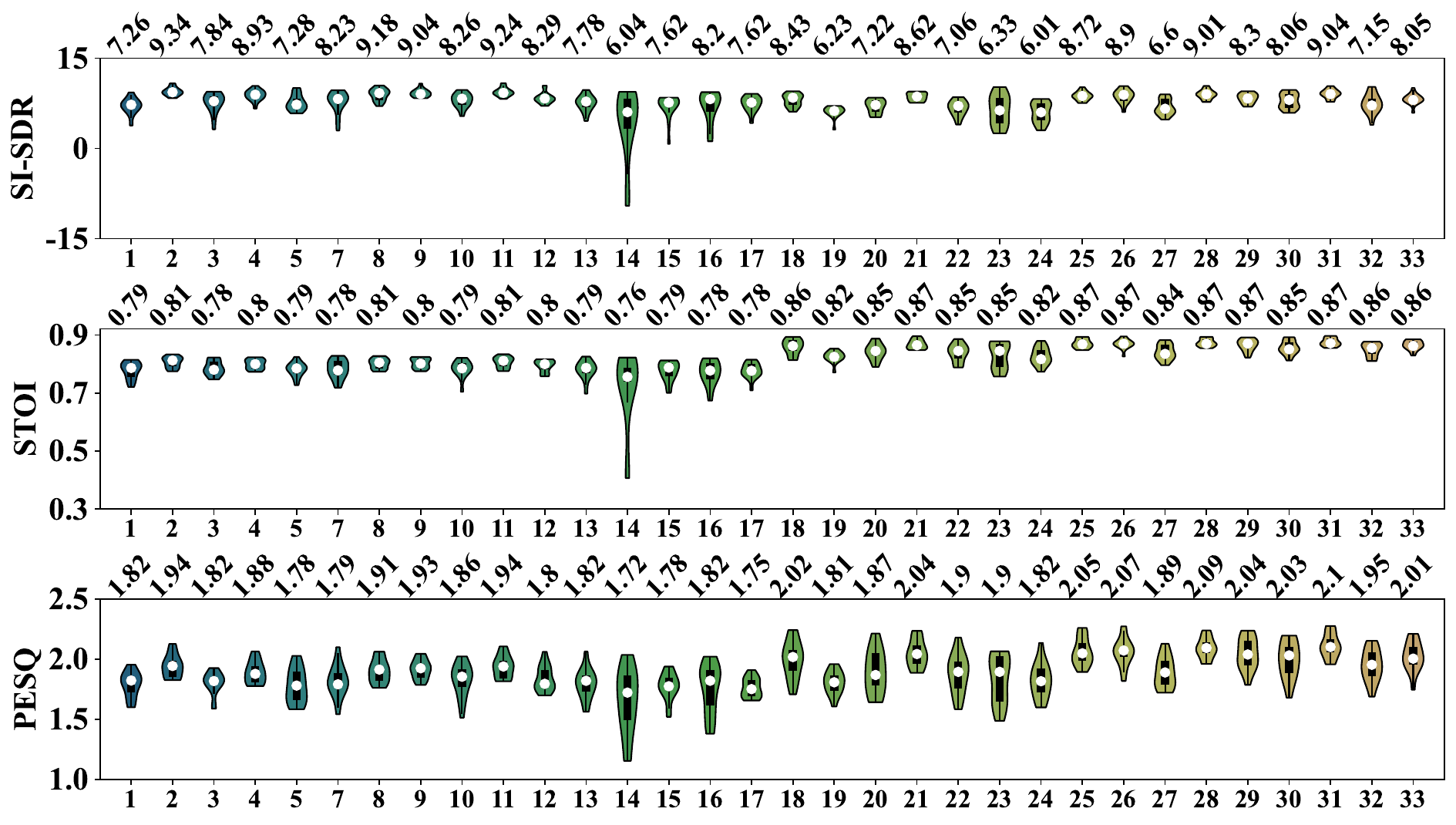}}
  \subfigure[Proposed BASEN]{
  \centering
    \includegraphics[width=0.48\textwidth]{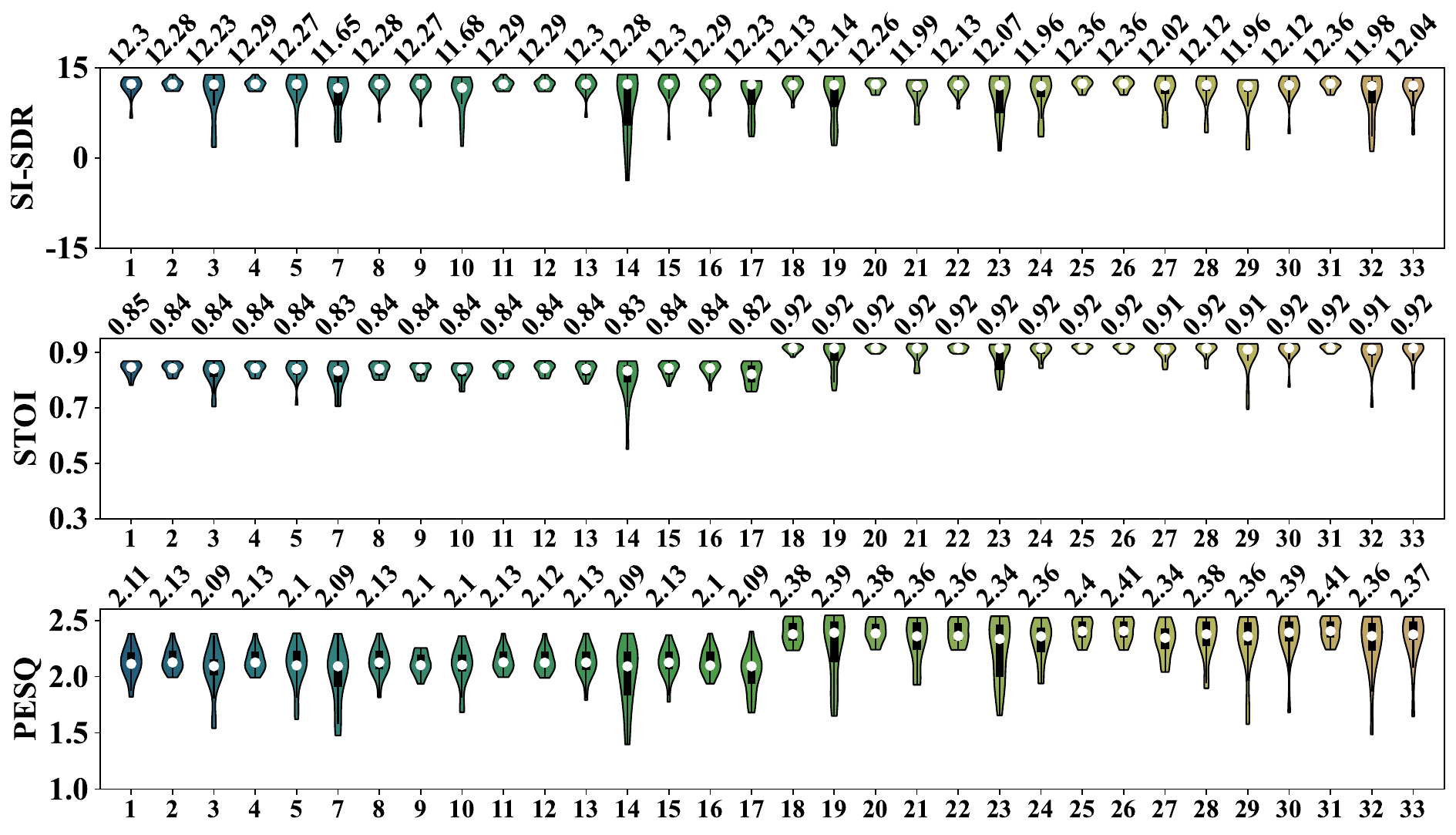}}
  \caption{The comparison of the proposed BASEN baseline with the UBESD method, where the median values are shown at the top of each sub-figure. The white dot in each plot shows the median. The black bar in the center of the violins shows the interquartile range (IQR). The thin black lines stretched from the bar show {\it first quartile $-1.5\times IQR$} and {\it third quartile} $+1.5\times IQR$, respectively.}
  \label{Figure3}
\end{figure*}

\subsection{Evaluation of the proposed baseline BASEN}
First, we show the effectiveness of the proposed baseline BASEN in comparison with  UBESD~\cite{hosseini_end--end_2022} and analyze the impact of each module.  Fig.~\ref{fig:self-comparison}(a) shows the performance of BASEN using the audio-only signal, simple concatenation and/or CMCA.  It is clear that without prior information of the attended speaker, the typical  Conv-TasNet (i.e., BASEN-EEG-prior) cannot improve the speech quality, since both speakers produce speech signals and no background noises are present in the considered setting. Given the attended speaker, Conv-TasNet becomes equivalent to the proposed BASEN without EEG trials but with prior information, which can largely improve the performance.  This shows that the efficacy of Conv-TasNet~\cite{luo2019conv} heavily depends on the  attended speaker information.  Including the EEG branch and simply concatenating the audio and EEG embeddings can achieve a comparable performance as the ideal Conv-TasNet. Applying the proposed CMCA module for feature fusion further improves the performance, particularly in terms of PESQ.

In order to find out the most appropriate number of cross-attention layers in the CMCA module, we  evaluate the impact of the layer number on the performance in Fig.~\ref{fig:self-comparison}(b), where $N$ changes from 1 to 5.  It is clear the choice of $N$ = 3 returns the best SE performance in all metrics, which will thus be used in the sequel. {Note that including more layers will also increase the parameter amount of the overall BASEN model, e.g., from 0.57M for $N$ = 1 to 0.64M for $N$ = 3.}

Further, we compare the proposed BASEN with the best published method on the same dataset, i.e., UBESD~\cite{hosseini_end--end_2022}. The obtained SI-SDR, PESQ and STOI across testing subjects are summarized in Fig.~\ref{Figure3}. We can clearly see that the proposed BASEN outperforms the UBESD in all metrics and over all subjects. More importantly, the performance of BASEN is more subject invariant, as that of UBESD varies more seriously across subjects, meaning that BASEN is more robust against listening dynamics. This also shows that the proposed CMCA module is more effective for feature fusion than the FiLM strategy in~\cite{hosseini_end--end_2022}. {Note that  UBESD  also has a larger parameter amount (around 1.84M) due to the fact that the adopted FiLM has four convolutions and each has more channels.}

 { 
\subsection{Subject-Trial Independent Evaluation of BASEN}

\begin{figure}[!t]
\centering
    \includegraphics[width=0.4\textwidth]{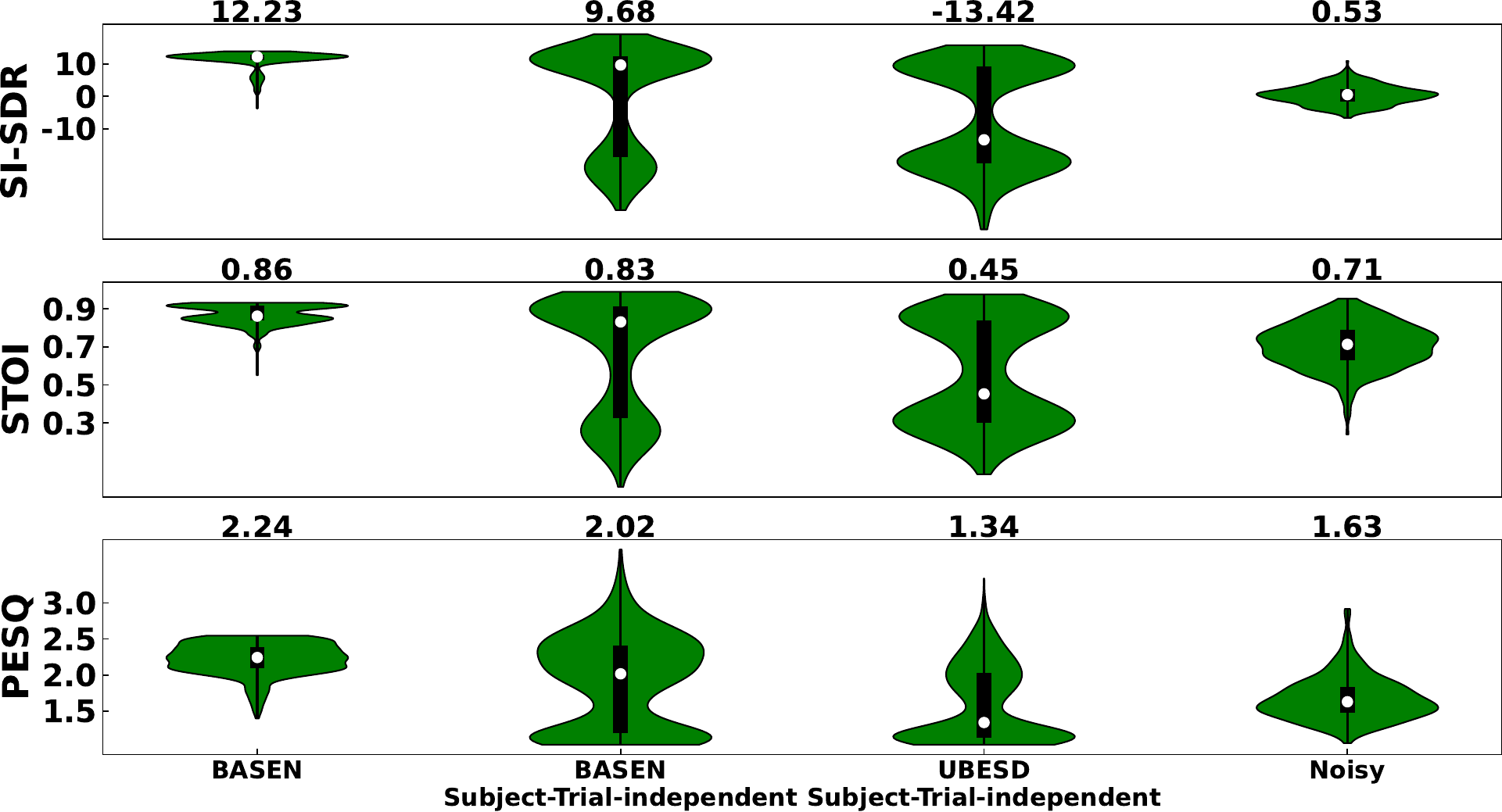}
  \caption{Subject-trial independent experimental comparison, where `BASEN' shows the performance obtained using the data pre-processing in Sec.~\ref{sec:preprocessing}.}
  \label{fig:ind}
\end{figure}

As DNNs generally have the ability of discovering and learning distinctive features from training samples, the proposed BASEN might recognize to which trial a test EEG segment belongs if the train and test segments are close in time in the EEG recordings according to the dataset split in Sec.~\ref{sec:preprocessing}. That is, training and testing on the same speakers in the mixture might be biased{~\cite{li2020perils,rotaru2024we}}, as the model can learn the target voices and extract them without the need to decode the auditory attention clue from EEG signals.
In order to avoid that the network learns to recognize the EEG of a subject to select the attended speaker, we reconsider the data split to evaluate the efficacy of the proposed BASEN in a joint  subject  and trial independent case. Due to the limited dataset size,  we randomly select 5 subjects from all subjects as the test group and 2 subjects as the validation group, and all the remaining subjects are used for training (roughly 9.5-hours audio that are enough to train Conv-TasNet compared to the model size). All trials of each subject are split similarly, i.e., 23 trials for training, 2 trials for validation and 5 trials for testing. This ensures that both the testing subjects and trials are unseen in either training or validation subset. Note that this splitting fashion roughly reduces the valid data amount of 15\% for training and 80\% for testing, respectively.
\begin{figure*}[!t]
  \centering
  \subfigure[The performance for ResGS]{
  \centering
    \includegraphics[width=0.42\textwidth]{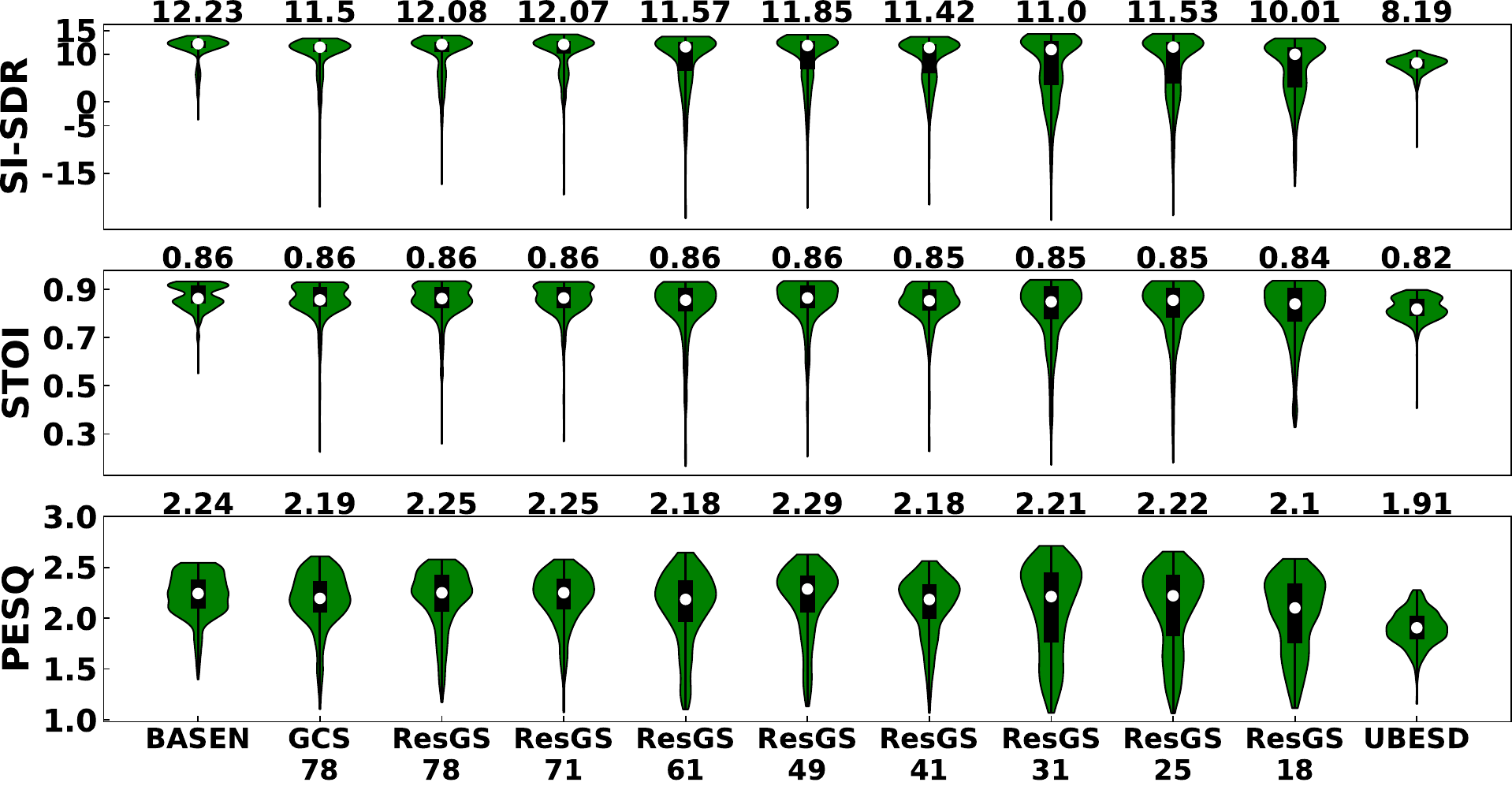}}
  \subfigure[The performance for ConvRS]{
  \centering
    \includegraphics[width=0.42\textwidth]{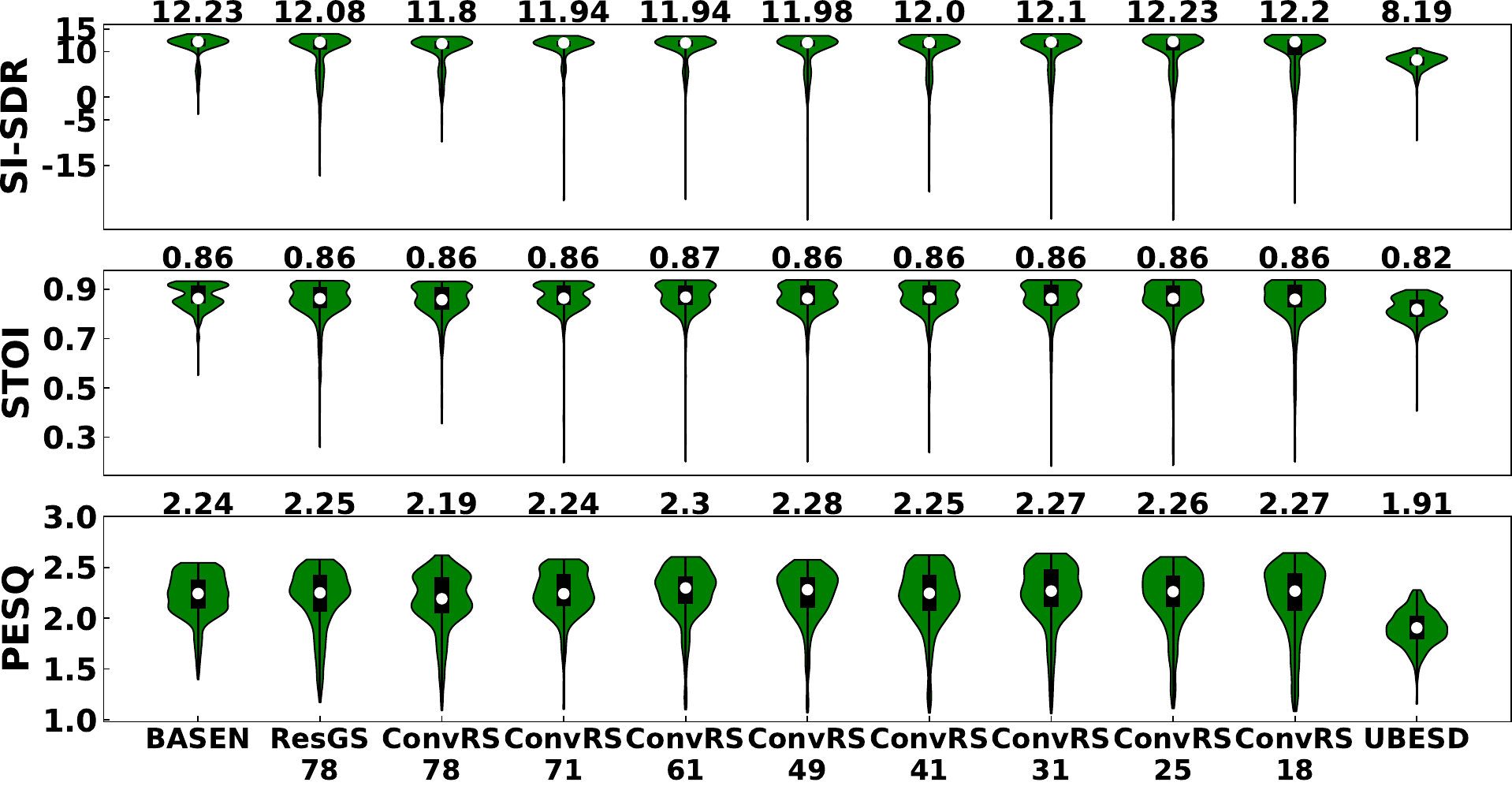}}
  \caption{The performance of the proposed EEG channel selection methods in terms of SI-SDR, PESQ and STOI (left: ResGS, right: ConvRS).}
  \vspace{-0.2cm}
  \label{fig:selection}
\end{figure*}

The obtained  results are shown in Fig.~\ref{fig:ind}, where we also include the performance of BASEN as the upper bound from Fig.~\ref{fig:self-comparison}, UBESD under the same condition and the noisy indicators for comparison. Note that `Noisy' metrics are computed using the new test subset, which thus slightly differ from those in Fig.~\ref{fig:self-comparison}. It is clear that in this thoroughly independent case the proposed BASEN can still significantly improve the perceptual quality of the target speaker and perform much better than the SOTA UBESD method~\cite{hosseini_end--end_2022}.  It seems that UBESD is not trained successfully in this subject-trial independent case, as the model size ($\approx$1.84M) is much larger compared to the training data. We notice that this new data splitting causes a larger variance in performance, which is mainly due to the decreased data amounts, particularly for testing. As the average performance of the upper-bound BASEN does not exceed the independent counterpart too much, the data split in Sec.~\ref{sec:preprocessing} would not lead to a serious bias and the model cannot learn the target voice only depending on the audio modality in this listening experiment. EEG signals are still necessary in this case. This verifies the rationality of the data split in Sec.~\ref{sec:preprocessing}, which will therefore still be used for further performance analysis and fair comparison with UBESD~\cite{hosseini_end--end_2022} in the sequel.
}

\subsection{Evaluation of ResGS-based EEG Channel Selection}

Second, we validate the proposed ResGS-based EEG channel selection method in comparison with the BASEN baseline and UBESD~\cite{hosseini_end--end_2022}, which involve all 128 EEG channels, as well as the classic GCS~\cite{strypsteen2021end} with 78 channels (GCS-78). {Due to the insufficient training stablity of the classic GCS, we use the selection layer of GCS after training and freeze it to train a new BASEN to evaluate the GCS-based SE performance.} From Fig.~\ref{fig:selection}(a), we can clearly see that with the decrease in the channel number (ranging from 128 to 18), the performance of the proposed method slightly changes, which is very close to the optimal BASEN performance in all three metrics and is { comparable with GCS-78}. This indicates that  a small group of EEG channels contributes largely to the EEG-audio SE task, and many EEG channels are marginal. This also shows that our ResGS method can select the important channels effectively and successfully resolve the training issue in the original GCS (which is more awkward to be combined with the back-end BASEN for SE). It is interesting that the proposed ResGS with 18 channels still outperforms UBESD in all three metrics. It is noticeable that for the proposed method the SI-SDR monotonically changes in terms of the number of the selected EEG channels due to the considered loss function, while this does not hold for STOI or PESQ, as reducing the channel number can even improve the perceptual quality in some cases. This implies that some EEG recordings might be  even noisy and the exclusion of these channels might be beneficial for the target speech perception.
\begin{figure}[!t]
\centering
    \includegraphics[width=0.4\textwidth]{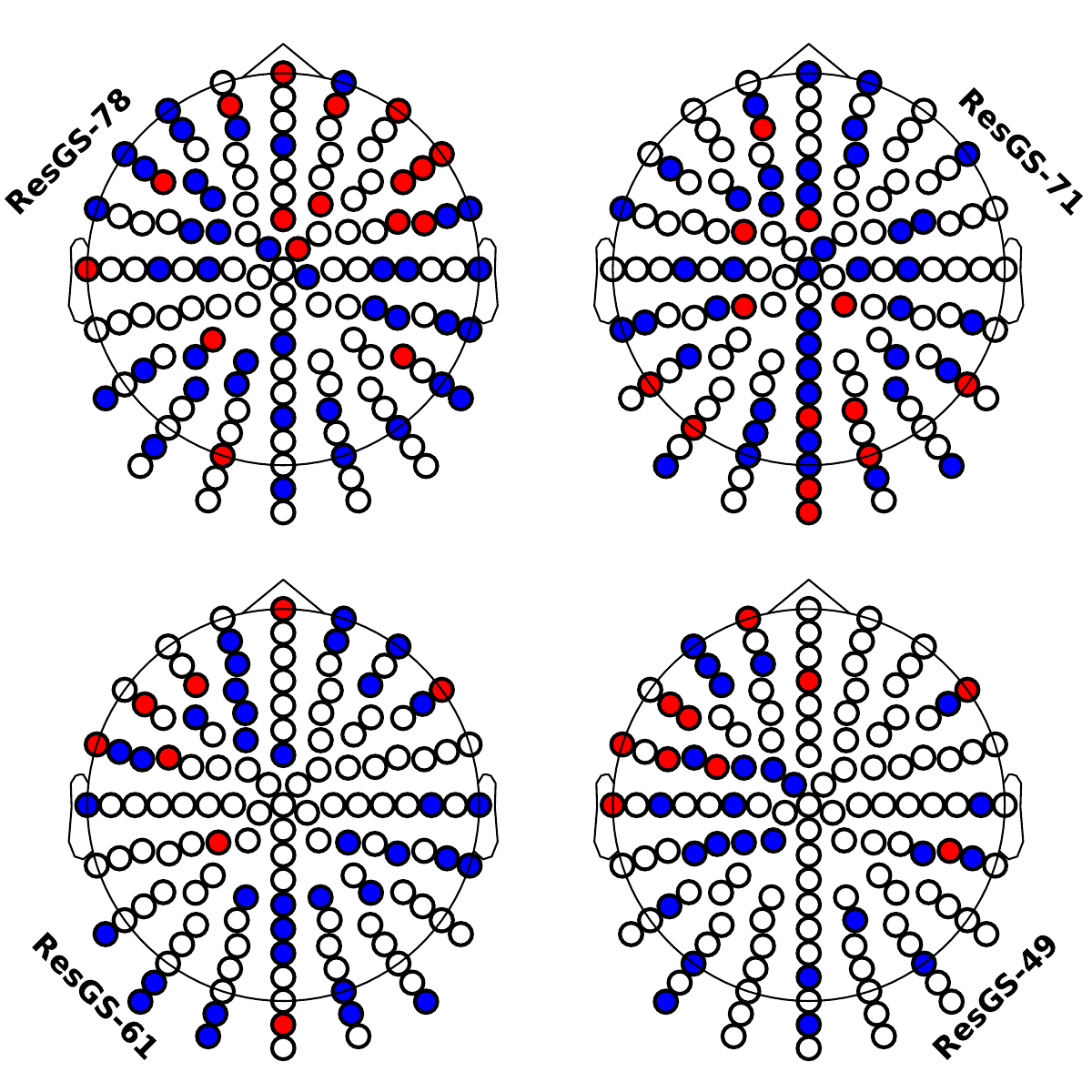}
  \caption{Visualization of the selected EEG channels by ResGS, where the blue and red dots denote unique and duplicated selections, respectively, and the remaining circles are unselected eletrodes.}
  \label{fig:v1}
\end{figure}

In order to find the informative EEG channels by ResGS, we visualize some selection examples (e.g., 49, 61, 71, 78) in Fig.~\ref{fig:v1}, where the blue and red dots are unique and duplicated selected channels, respectively. {We can see that the channels around the ears, left and right temporal cortex and the forehead are more likely to be chosen.  This is  consistent with the conclusion in~\cite{preisig2021selective}. In ResGS-49, it is clear that the channels on the left brain are more important. This is related to the fact that right-handed language processing is mainly in the left brain.} The emergence of duplicated channels is caused by the inherent problems in GCS, where the regularization can only alleviate this effect but can not avoid completely especially in the case of a large channel number~\cite{strypsteen2021end}, although the proposed ResGS can resolve the issue of training instability.
\begin{figure*}[!t]
  \centering
  \subfigure[Full-channel UBESD]{
  \centering
    \includegraphics[width=0.48\textwidth]{fig/ubesd.pdf}}
  \subfigure[ResGS with 49 channels (ResGS-49)]{
  \centering
    \includegraphics[width=0.48\textwidth]{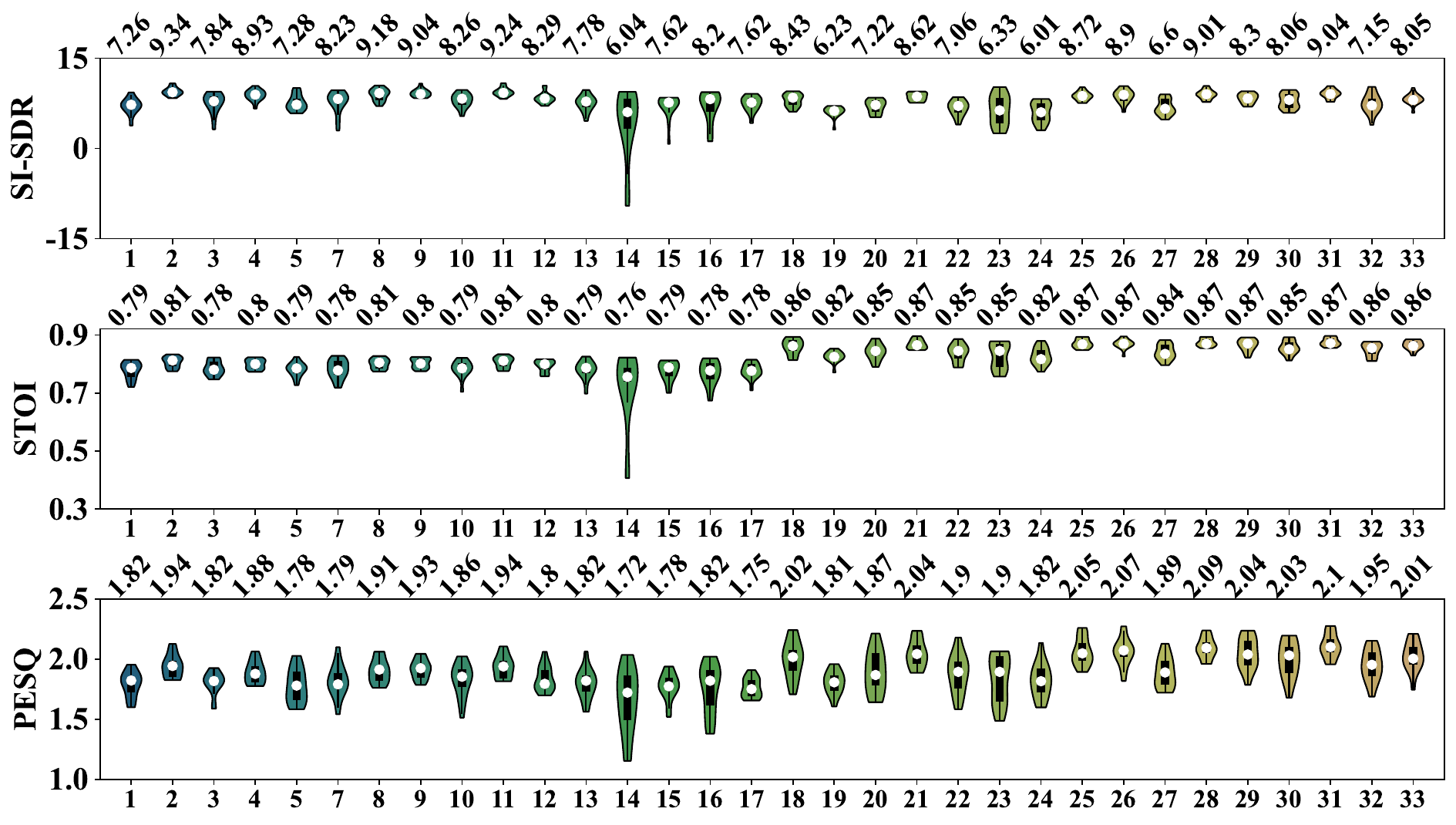}}
\subfigure[ResGS with 18 channels (ResGS-18)]{
  \centering
    \includegraphics[width=0.48\textwidth]{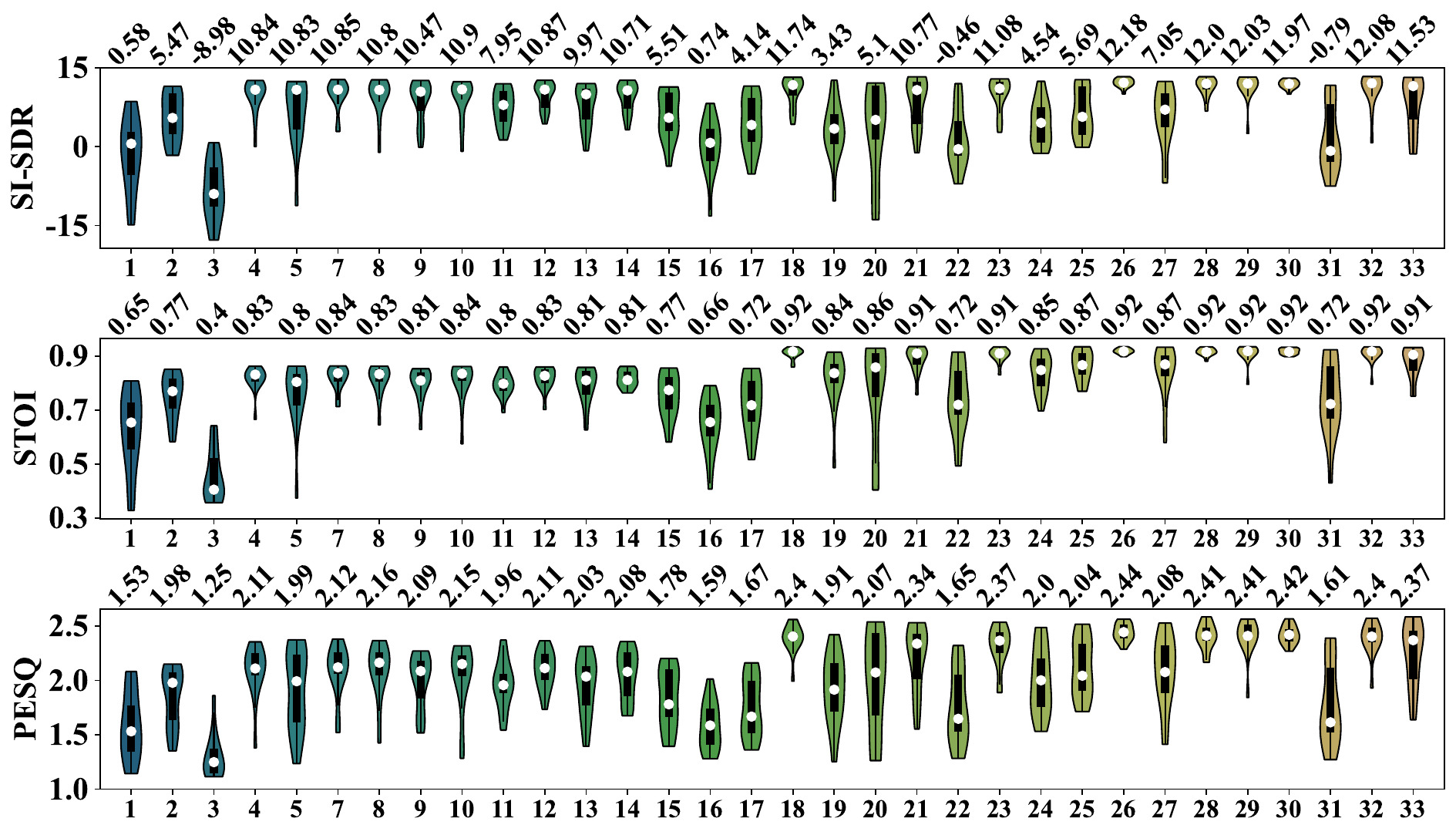}}
  \subfigure[ConvRS with 18 channels (ConvRS-18)]{
  \centering
    \includegraphics[width=0.48\textwidth]{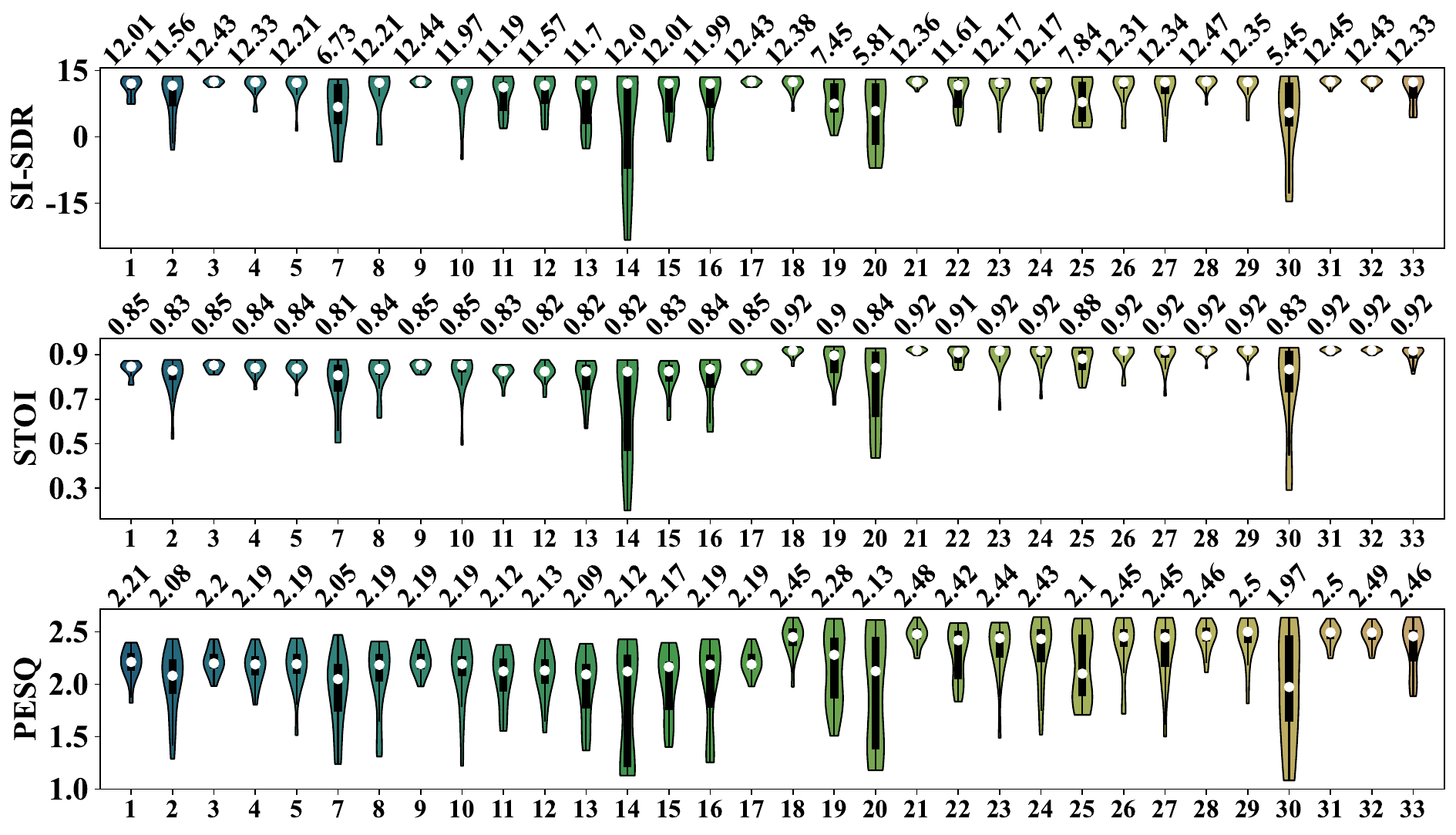}}
  \caption{The performance comparison of the full-channel UBESD, ResGS with 18 and 49 channels and ConvRS with 18 channels across subjects.}
  \label{fig:select-sub}
\end{figure*}

\subsection{Evaluation of ConvRS-based EEG Channel Selection}
Third, we show the effectiveness of the proposed ConvRS-based EEG channel selection approach in comparison with the full-channel inclusions for BASEN and UBESD and the ResGS with 78 channels. The sparse selections of ConvRS are obtained by adapting the sparsity regularizer $\gamma$ picked from \{0, 0.05, 0.1, 0.15, 0.2, 0.25, 0.3, 0.35\}, corresponding to the selected subsets of \{78, 71, 61, 49, 41, 31, 25, 18\} channels, respectively. The results are shown in Fig.~\ref{fig:selection}(b), from which we can observe that the SI-SDR increases slightly as the channel number decreases except for 18 channels. The STOI almost keeps unchanged in terms of the selected EEG subsets. The PESQ can be improved by removing non-informative channels compared to the typical BASEN.
This means that many EEG channels might contain too much noise and are thus useless to  target speaker extraction, and it is even possible that some channels have a negative contribution. The removal of irrelevant EEG channels by the proposed ConvRS method is beneficial for target speech perception, which can at least achieve a comparable SE performance with the full inclusion. Comparing with the results of ResGS in Fig.~\ref{fig:selection}(a), given the same number of selected channels the ConvRS method outperforms ResGS when the selected subset is sparser, indicating a stronger robustness against sparsity.

\begin{figure}[!t]
\centering
    \includegraphics[width=0.46\textwidth]{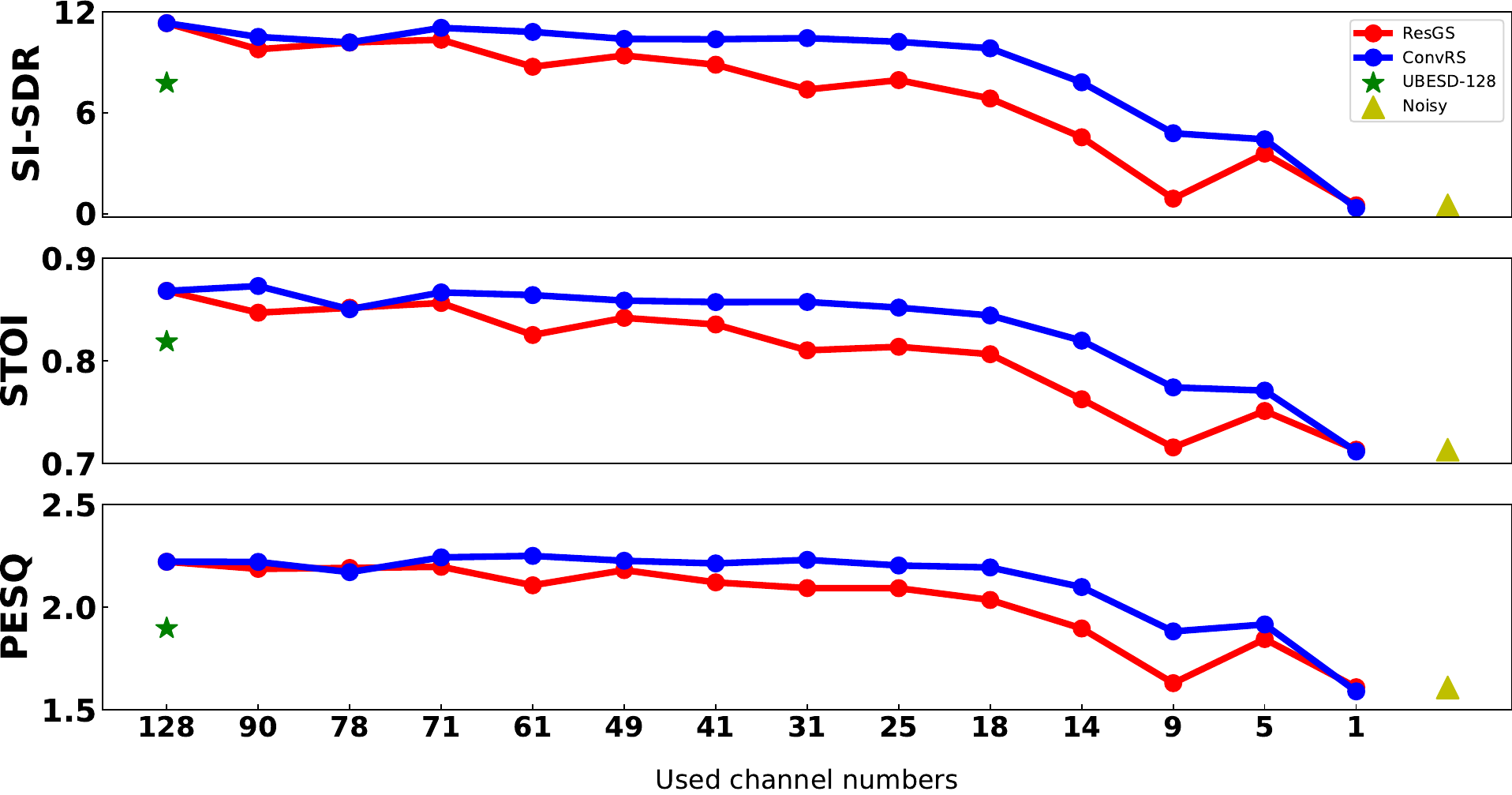}
  \caption{Performance change in terms of the number of selected channels.}
  \label{fig:conv}
\end{figure}

{As the attention decoding performance only starts to suffer when the number of EEG channels is 10 or lower, it is interesting to see how the channel selection mechanism works on extremely small channels subsets. In addition to the preset number of selected channels in Fig.~\ref{fig:selection}, we show the average performance of ResGS and ConvRS in comparison to the full-channel UBESD in Fig.~\ref{fig:conv} for completeness. It is clear that the performance generally decreases in terms of the channel amount, and given the same number of selected channels, the proposed ConvRS achieves a better performance than ResGS.  In the case of more than 10 selected channels, the proposed methods can consistently outperform the full-channel UBESD. If the preset channel number is extremely small ($\leq$ 9), the performance varies significantly, which however is still better than the noisy input mixture.}

\begin{figure*}[!t]
  \centering
    \includegraphics[width=0.85\textwidth]{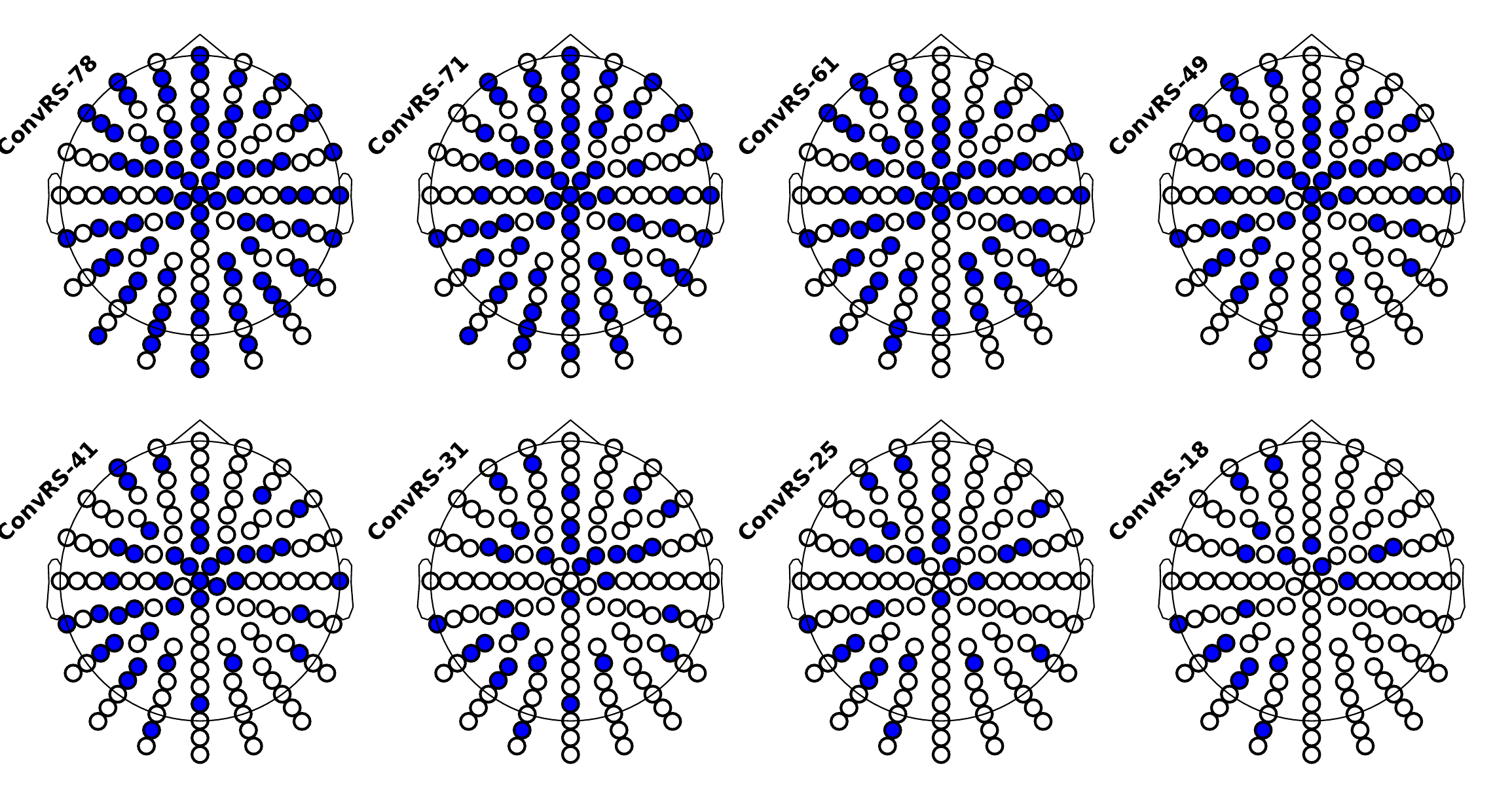}
  \caption{Visualization of the selected EEG channels by ConvRS, where the blue dots denote  unique selections and no duplicated selection occurs.}
  \label{fig:visual}
\end{figure*}

Similarly, we visualize the selected channels (e.g., 18, 25, 31, 41, 49, 61, 71, 78) of ConvRS by changing different $\gamma$-values in Fig.~\ref{fig:visual}.  {Note that the selection solutions of ConvRS is not input-dependent on the considered dataset as stated in Sec.~\ref{sec:progress_training}, which only depends on the user-defined controller $\gamma$. The selected electrodes are resolved by simply rounding the selection vector $\mathbf{s}$ after training.} Compared to the selections of ResGS in Fig.~\ref{fig:v1}, the proposed ConvRS method contains no duplicated channels. Therefore, given the same number of channels to be chosen, the resulting subset of ConvRS will incorporate more electrodes than that of ResGS, but all selected channels are unique. We can also see that the channels around the ears, left and right temporal cortex and the forehead are more likely to be chosen. {This observation is consistent with the findings by ResGS in Fig.~\ref{fig:v1}. For example,  in ConvRS-18 the channels on the left brain are clearly more important, which is similar to the selected channels of ResGS-49.}

Finally, we evaluate the performance variation across subjects in Fig.~\ref{fig:select-sub}, where the performance of ConvRS-18, ResGS-18 and ResGS-49 in comparison with the full-channel UBSED is shown  in terms of SI-SDR, STOI and PESQ. For most subjects, given 18 selected channels, ConvRS outperforms  ResGS and shows a better robustness against the subject variation. This is due to the fact that ResGS might result in duplicated selections and all selections of ConvRS are unique. Compared to UBESD, it is clear that for most subjects ConvRS-18 and ResGS-49 can achieve a better performance. It is interesting to note that using more EEG channels could improve the performance to be more subject invariant, e.g., the metrics of UBSED and ResGS-49 change more slowly than the other two cases.  Also, we can observe that some subject-specific performance is relatively lower compared to other subjects. This is caused by the subject variation, since the subjects' concentration levels are different during data collection, and neither the proposed ResGS nor ConvRS has taken the concentration level for designing a subject-adaptive channel selection into account.

\section{Conclusion and Limitation}\label{sec:conclusion}
In this paper, we investigated the exploitation of sparse EEG channel selections for the time-domain brain-assisted SE task. First of all, we built the BASEN baseline model using EEG and audio signals, which shows that without any prior knowledge on  the listener attention, the EEG signals implicitly contain this clue and the inclusion of EEG signals for blind speech separation is thus helpful. Compared to UBESD, the superiority of the proposed BASEN lies in the  Conv-Tasnet backbone and the designed CMCA module for deep bi-modal feature fusion. As in practice the design of an EEG cap with many electrodes would cause several problems, a sparse EEG channel distribution that can guarantee a comparable performance is thus more preferable. We then considered two sparse channel selection methods for the proposed BASEN, which are called ResGS and ConvRS, respectively. They are respectively dedicated to solving the training instability of existing channel selection approaches (but for other tasks) and the duplicated selections. Experiments on a public dataset showed the efficacy of the combination of BASEN with channel selectors. 

Using the informative EEG channel subsets (even a subset of 18 channels) for BASEN does not decrease the average performance too much with respect to the full-channel incorporation, which can even improve the performance in some cases. This validates that multichannel EEG measurements are highly correlated, some channels are irrelevant or marginal to target speaker extraction. The exclusion of these channels has a negligible impact, which however has a rather important practical value of saving e.g.,  hardware cost, setup time, algorithmic complexity. Besides,  {channel selection and visualization can help to find out the brain areas that are relevant to spatial speech perception. The obtained informative channels of both ResGS and ConvRS are consistent with the findings in the context of binaural integration~\cite{preisig2021selective}.}

\begin{figure}[!t]
\centering
    \includegraphics[width=0.25\textwidth]{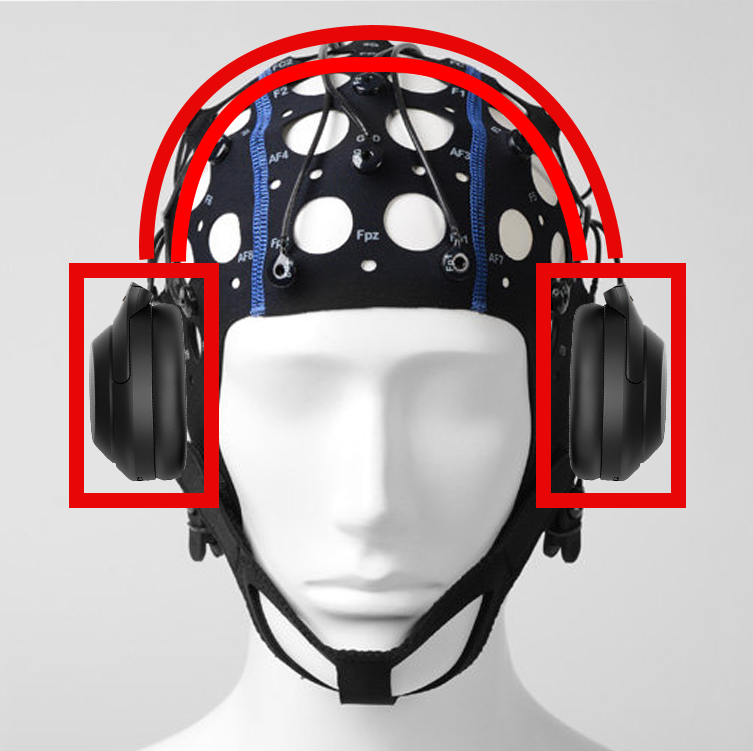}
  \caption{An example of topology-constrained integration of hearing-aid and EEG devices, where only the red region is allowed to place EEG electrodes.}
  \label{fig:top}
\end{figure}
{The limitation of the proposed sparsity-driven SE methods is threefold: 1) Channel selection would cause a larger performance variation across subjects. It is desired to achieve a more stable performance over testing subjects. 2) Due to the limited data amount, given a user-defined sparsity controller,  the obtained selection solution seems input-independent. It is smarter to achieve a selection adaptive to listening preferences and brain structures of subjects. These two points can be considered as subject-adaptive EEG channel distribution in the future work. 3) The topology constraint is also an important concern that has to be taken into account when combining hearing-aid devices and EEG electrodes for hardware integration, as in this context the EEG electrodes are only allowed to be placed in the restricted area on the cap, e.g., the red area in Fig.~\ref{fig:top}. This can be done by a two-step selection strategy, e.g., using the topology constraint to perform an initial electrode selection and then applying the proposed methods to refine the solution with respect to the initially obtained subset. The integration of EEG and assistive hearing devices should also consider using EEG signals to assist joint binaural speech separation and spatial cues preservation, where the latter is vital to the localization of directional stereo sounds. 

To our knowledge, this work is the first attempt of incorporating sparse EEG channel selection for brain-assisted speech processing.  Indeed, it still belongs to the field of monaural SE, and the efficacy has to be further validated in a binaural setting. For this, we will release a dataset consisting of matched (synchronized) binaural audio recordings and EEG measurements of listeners. As the focus of this work is on SE, the adopted performance metrics are all related to SE. Since the channel selection affects the SE performance via AAD indirectly, it would be interesting to analyze how it impacts AAD. This will be shown in an independent short report.
}
 
\section*{Acknowledgments}
The authors would like to thank the anonymous reviewers for their helpful remarks and constructive suggestions. Also thanks to Richard C. Hendriks (at Faculty of EEMCS, Delft University of Technology) for warm discussions. The complete implementation code for this paper will be available at {\url{https://github.com/jzhangU/Basen}}.

\bibliographystyle{IEEEtran}
\bibliography{ref}

\vfill

\end{document}